\documentclass[useAMS,usenatbib]{mn2e}
\usepackage{longtable}
\usepackage{graphicx}
\usepackage{natbib}
\usepackage{lscape}
\usepackage{rotating}

\def \nuprocess{$\nu$-process}

\def \kms{km s$^{-1}$}
\def \nodata{. . .}
\def \degree{$^{\circ}$}
\begin{document}
\title{A Search for Boron in Damped Ly$\alpha$ Systems}
\author[Berg et al.] {Trystyn A. M. Berg$^1$,  Sara L. Ellison$^1$, Kim A. Venn$^1$, J. Xavier Prochaska$^2$  \\
$^1$ Department of Physics and Astronomy, University of Victoria, Victoria, British Columbia, V8W 2Y2, Canada.\\
$^2$ Department of Astronomy and Astrophysics, University of California, Santa Cruz, Santa Cruz, CA, 95064, USA.}
\maketitle
\begin{abstract}

We present the first systematic study of boron beyond the Local
Group. This analysis is performed on a sample of 30 damped Ly$\alpha$ systems (DLAs) with strong metal-lines, which are expected to trace the interstellar medium of high $z$ galaxies.  We report on two boron detections at $>3\sigma$ significance; one new detection and one confirmation.  The ratios of B/O and, for the
first time, B/S are compared with previous stellar and interstellar
measurements in the Milky Way and Small Magellanic Cloud.  The novel
comparison with sulphur, which tracks oxygen's abundance, alleviates the uncertainty associated with stellar oxygen measurements.  For both detections, the inferred B/S
ratio is in excess of the prediction of primary boron production from spallation processes. Possible sources of contamination are discussed, as well as physical effects that could impact the observed ratios. However taken at face value, the implication of these measurements suggest potentially higher cosmic ray fluxes in DLAs. The prospects
for future boron detections in other high redshift DLAs to confirm our results is also discussed.

\end{abstract}
\section{Introduction}

Although most elements are primarily formed through stellar or
Big Bang nucleosynthesis, beryllium and boron are notable exceptions. \cite{Reeves70}
first described Be and B formation through the spallation of galactic
cosmic rays (GCRs) with CNO nuclei. The original picture of Be and B
production from spallation involved interstellar protons and alpha
particles that are accelerated by supernovae and which subsequently collide
with interstellar CNO. The production of B and Be through this
\emph{forward} mechanism (also known as \emph{direct} spallation)
depends on both the rate of GCR production (i.e. supernova rate) and
on the metallicity of the interstellar medium (ISM).  Forward
spallation is therefore considered to be a \emph{secondary} process
with a predicted dependence of Be and B $\propto$[CNO]$^{2}$ (i.e.
$m=2$ where $m$ is the slope of the logarithmic dependence between
Be and B with CNO).

However, the simple model of forward spallation is in conflict with
the metallicity-independent production of the light elements which is
observed in halo stars \citep[e.g.][]{Duncan92,Boesgaard99Be}. Models
of B and Be production have therefore tried to identify \emph{primary}
mechanisms (Be, B $\propto$[CNO], $m=1$) such as the \nuprocess{} in
stars \citep{Woosley90}, or spallation in which the GCRs always have
the same CNO content \citep{Duncan92}. The latter class of processes
is referred to as \emph{reverse} spallation and generally entails the
acceleration of CNO nuclei which then collide with ambient protons or
alpha particles.  Various mechanisms have been suggested to accomplish
reverse spallation, including supernovae accelerating either 1) their
own ejecta, 2) locally enriched superbubble material or 3)
wind-enriched material around massive rotating stars
\citep[see][for a review]{Prantzos12}.  In reality, there may be multiple
processes that contribute to Be and B production, where an intuitive
combination may be one where reverse spallation (with interstellar
protons and alpha particles as the targets) dominates in a metal-poor
ISM.  As the ISM enriches and more CNO targets accumulate, forward
spallation can become more effective.  Quantifying the relative
contributions of different processes requires detailed chemical
modeling which can simultaneously account for the observed abundances
of B, Be and Li (which can also be produced through spallation), as
well as isotopic ratios \citep{Fields00,Prantzos12}.

Meanwhile, observational studies of boron are quite limited. The strongest 
boron transitions required for determining
stellar abundances are situated in the ultra-violet (UV) and have very
weak oscillator strengths.  Furthermore, boron astration due to
rotational mixing in stars also impacts measurements
\citep{Venn02,Mendel06}.

In addition to the observational challenge of measuring boron, a
further complication for the interpretation of boron (and Be)
abundances is the comparison with the oxygen abundance
\citep[e.g.,][]{Fields00}.  Despite its high cosmic abundance and
variety of spectral features available for measurement, determining
accurate O/H abundances has been a subject of much controversy
\citep[e.g.,][and references therein]{Boesgaard99O,Israelian01}.  In
brief, there are four different features that are commonly used for
determining oxygen abundances (the [OI] $\lambda \lambda$6300, 6363
\AA{} forbidden lines, near infra-red (IR) OH vibration-rotation
$\lambda \lambda$1.6, 3.4 $\mu m$ lines, the UV OH $\lambda$ 
3100-3200\AA{} lines, and the OI triplet at $\lambda$ 7771-7775 \AA{})
 yet they result in internally inconsistent abundances. Specifically, the UV OH lines tend
to find a rising [O/Fe] at low [Fe/H], in contrast with an [O/Fe]
plateau from [OI] measurements \citep[e.g.][]{Kraft92,Carretta00}.
It is clear that the uncertainty in oxygen abundances complicates both the direct
comparisons of Be and B with O, but also hinders observational studies
that rely on converting Fe to O via calibrated relative abundances.
For example, the boron study of \cite{Smith01} attempted to derive
oxygen abundances in halo stars from Fe measurements taken from
\cite{Duncan97} and \cite{GarciaLopez98}. To convert from Fe to O,
three different models were applied to describe [O/Fe] as a function
of [Fe/H], representing the uncertainty in the conversion.  Depending
on the choice of conversion, they determined a dependence of B on O
that ranged from purely primary ($m$=0.92, 1.05), to a mix of primary
and secondary processes ($m=$1.44).  An accurate interpretation of Be
and B production mechanisms clearly requires a reliable oxygen (or
proxy) abundance.

Despite these challenges, observations of Be and B in stars have been
used to infer that their production is a combination of
both primary and secondary processes. \cite{Smith01} determined oxygen
abundances from the weak [O I] lines for 13 F and G field stars for
which \cite{Cunha00} derived the boron abundances. The best-fit slope
of the B-O relation was found to be $m=1.39$, suggesting a combination
of primary and secondary processes. This agrees with an analysis of
beryllium done by \cite{Rich09}, who observed 24 stars and compared
the best-fit double power model ([O/H]$>-1.8$ and [O/H]$<-1.4$; $m=1.59$,
0.74 respectively) and a single line best fit ($m=1.21$) of their Be-O
relation. They found that there appeared to be a two-component trend
with a break point at [O/H]$\sim-1.6$ ([Fe/H]$\sim-2.2$), 
indicating a transition from primary to secondary processes.

\textit{Interstellar} boron can also contribute to our understanding of its
origin, although relatively few observational studies exist.  The
first observation of interstellar boron was made by \cite{Meneguzzi80}
using the Copernicus satellite.  However, it was the launch of the
Hubble Space Telescope (HST) that allowed real progress in this area
with observations made first with the Goddard High Resolution
Spectrograph (GHRS) \citep{Federman93,Jura96}, and later with the
Space Telescope Imaging Spectrograph (STIS) \citep{Howk00}.
Importantly, \cite{Howk00} found that boron can be significantly
depleted onto dust in the ISM, leading to potential under-estimates of
its actual abundance.  They found that boron is more easily depleted
in the cold diffuse ISM, and demonstrated an anti-correlation of the
boron abundance with hydrogen gas density.  The most recent
measurements of interstellar boron have been made by \cite{Ritchey11}
who added a further 56 Galactic sight-lines observed with STIS.  Based
on their sight-lines through warm, low density gas (assumed to be
relatively undepleted) \cite{Ritchey11} determine an ISM abundance of
B/H $=2.4\pm0.6 \times 10^{-10}$, agreeing with the results from
undepleted B-type stars \citep{Venn02} and previous ISM studies
\citep{Howk00}.

Beyond the stellar and interstellar measurements in the Milky Way,
there has been only one extra-Galactic study of boron.
\cite{Brooks02} observed two B-type stars in the Small Magellanic
Cloud (SMC) in an attempt to compare boron production relative to the
Milky Way. The two boron upper limits presented by Brooks et al. (2002) lie below the expectation for primary boron production, but are
consistent with a secondary production mechanism.  However,
Brooks et al. (2002) acknowledge that their measurements do
not account for depletion of boron due to rotational mixing,
hence they suggest that their limits do not necessarily rule out primary
production.

The interpretation of boron abundances in other galaxies depends on both
the cosmic ray flux (CRF) and the abundance of spallation targets. The
SMC has an oxygen abundance around one quarter of that in the solar
neighbourhood \citep{Korn00,Salmon12}, and 15\% of its current CRF
\citep[based on Fermi/Large Area Telescope (LAT) gamma ray observations of the SMC,
see][]{Abdo10, Sreekumar93}.  \cite{Brooks02} point out that the lower
oxygen abundance and cosmic ray flux in the SMC might be expected to
have a compound effect that leads to a production of boron of
that is $\sim$1/20 lower than the solar neighbourhood for secondary boron
production, but also discuss the effect of higher past star formation
rates and CRF confinement times.  The discussion in \cite{Brooks02}
reveals that high boron abundances that exceed the primary dependence
(i.e. [B/H] $>$ [O/H]) can indicate high cosmic ray fluxes, which in turn
may be related to high rates of star formation.  The study of boron in other 
galaxies therefore provides a novel approach to constraining their star formation histories.

To study boron in other galaxies, and in particular, to push to higher
redshifts, damped Lyman alpha systems (DLAs) offer a plausible
prospect. DLAs are quasar absorption line systems defined to have a
column density of HI of N(HI)$\geq 2 \times 10^{20}$ atoms cm$^{-2}$,
and have been widely used to study the chemistry of high $z$ galaxies
\citep[e.g., see the review by][]{Wolfe05}. 
Conveniently for the present study, their cosmological distances provide the advantage of
observing the redshifted B II $\lambda1362$ line in the optical,
avoiding the use of space-based telescopes which currently limits
Galactic boron observations.

Due to the weakness of the B II $\lambda1362$ line, the most promising
first targets for boron observations in DLAs will be absorbers with
relatively strong metal lines.  A sample of such absorbers has been assembled by \citet{HerbertFort06}, who studied the so-called
metal-strong DLAs (MSDLAs), defined to have metal column densities log
N(ZnII) $\geq13.15$ and log N(SiII) $\geq15.95$. 
The possibility of boron detection
in MSDLAs has been demonstrated by a tentative detection in the
proto-typical MSDLA FJ0812+3208 ($z_{abs}$=2.626, log N(HI)=21.35)
by \cite{Prochaska03}.

In this paper, we present the first systematic search for boron in
DLAs.  We present a sample of 30 DLAs that have been selected as
promising targets for boron detection based on either the strength of
their metal lines, or overall metallicity.  A further novelty of our
study is that, for the first time in a boron study, we consider
sulphur as a proxy for oxygen to circumvent the problem with absolute
oxygen abundances described above.  The substitution of sulphur for
oxygen has been frequently used in DLAs, both for consistency in
comparisons with Galactic stellar data \citep{Nissen04,Nissen07} and
because the oscillator strengths of the OI transitions often
precludes an oxygen abundance determination \citep{Pettini02}. Although
both are alpha capture elements, it should be emphasized they do not have the same
nucleosynthetic origin.  Oxygen is primarily a product of helium and neon burning in massive stars, whereas sulphur is produced in oxygen burning and alpha-rich freezeout of core collapse supernovae \citep{Woosley95}. We therefore explicitly test the substitution of sulphur for oxygen. High
resolution echelle spectra have been obtained for each of the 30
target DLAs and measurements (or limits) made for the abundances of
oxygen, boron and sulphur.  The sample includes additional data for
FJ0812+3208, increasing the spectral signal-to-noise (S/N) and
permitting a re-analysis of this system.

\section{Targets of Observation }
\label{sec:targets}

\subsection{Sample Description}
\label{sec:sample}

Boron is an intrinsically rare element with a solar abundance one million times less than oxygen.  As a result of boron's low abundance and small oscillator strengths, it has intrinsically weak singly ionized lines.  We have therefore selected a sample of 30 DLAs that
is biased towards the most metal-rich DLAs, or those whose metal lines
are very strong due to high N(HI) column densities (see Table \ref{tab:MSDLAs}). Many of the 
objects were selected based on low-resolution spectra that indicated 
the possibility of high metallicities \citep[see][]{Kaplan10}. A further benefit of the typically high
metallicities of our sample is that the DLAs span the metallicity
range of the stars in which boron has been observed so far
in the Milky Way, facilitating a comparison between the two populations.

\begin{figure*}
\begin{center}
\includegraphics[width=\textwidth]{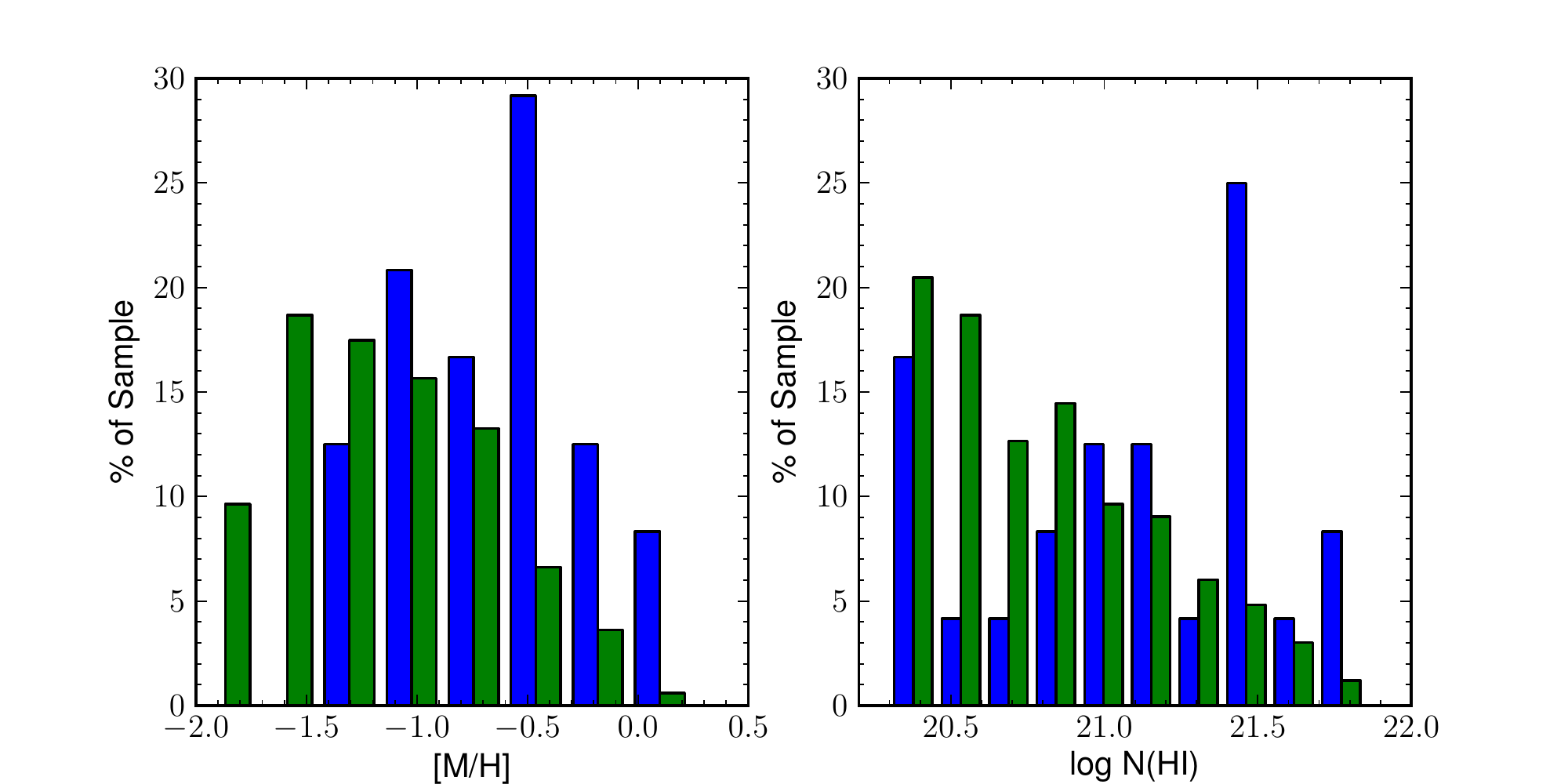}
\caption[Sample Histograms]{Distribution of metallicity ([M/H];
\emph{left}), and hydrogen column density(N(HI); \emph{right}). The literature sample from \cite{Rafelski12} is
plotted in green, and our metal-rich sample is plotted in blue. Note that absorbers without a measured hydrogen column density are excluded
from the histograms.}
\label{fig:hists}
\end{center}
\end{figure*}

The sample consists of 19 MSDLAs (selected purely on the basis of their metal
line strengths) and 11 DLAs with relatively high abundances. Metallicities for our sample were determined according to the scheme laid out by \cite{Rafelski12} where the assumed metallicity is the abundance of either sulphur, silicon, zinc, or iron\footnote{A $+0.3$ dex correction is included for iron to account for dust depletion.} (in order of decreasing preference). Metallicites of our sample along with the element used are tabulated in Table \ref{tab:BS}. The sample spans a wide range in N(HI) and specifically targets the metal-rich end of the literature sample presented by \cite{Rafelski12}. To emphasize this, Figure \ref{fig:hists} shows the distribution of metallicity ([M/H]; left), and hydrogen column density (N(HI); right) relative to the literature sample of 195 DLAs \citep[$z=0.091$--$4.743$][]{Rafelski12}. The left panel shows that higher metal contents are probed in our sample, along with a range of hydrogen column densities.

\subsection{Observations and Data Reduction}
All of the observations presented here have been made using the High Resolution Echelle Spectrometer (HIRES) spectrograph on the Keck I telescope.  Table 1 lists the QSOs that
have been observed, exposure time and the S/N ratio (per pixel)
near to the BII $\lambda$ 1362 \AA\ line (see Section \ref{sec:abund}).

The majority of data were taken with the HIRES spectrometer configured with the blue cross-disperser and the C1 decker giving a FWHM spectral resolution of $\approx 6 \, \rm km \, s^{-1}$.  The CCD was binned spatially by 2 and the spectra were optimally extracted to a fixed dispersion of 1.3\,km\,s$^{-1}$ per pixel.  Complete details on these observations will be presented in a future work (Berg et al., in prep.). The spectra were reduced using the \textsc{XIDL} code\footnote{http://www.ucolick.org/$\sim$xavier/IDL} following standard techniques. The
\textsc{HIRedux} software was used to reduce the spectra, and was fitted using
the \textsc{x\_continuum} routine which fits the continuum order by order
before converting into a 1D spectrum. 

\begin{table*}
\begin{center}
\caption{Target List}
\label{tab:MSDLAs}
\begin{tabular}{lcccccc}
\hline
Quasar & R.A. & Dec. & Magnitude (band) & Exposure Time (s) & S/N$_{1362}^{a}$ (pixel$^{-1}$)& MSDLA? \\
\hline
J0008$-$0958 & 00:08:15.3 & $-$09:58:54.0 & 18.4 (r) & 15029 &  11&Yes\\
J0058+0115 & 00:58:14.3 & +01:15:30.2 & 17.7 (r) & 14400 &  23&No\\
Q0201+36 & 02:04:55.6 & +36:49:18.0 & 17.5 (r) & 24980 & \nodata&No\\
Q0458$-$02 & 05:01:12.8 & $-$01:59:14.2 & 19.0 (R) & 28800 &  4&Yes\\
FJ0812+3208 & 08:12:40.7 & +32:08:08.6 & 17.5 (r) &40500  &45&Yes\\
J0927+1543 & 09:27:59.8 & +15:43:21.8 & 18.8 (r) & 12800  &4& Yes\\
J0927+5823 & 09:27:08.8 & +58:23:19.4 & 18.3 (r) & 21600  &14&No\\
J1010+0003 & 10:10:18.2 & +00:03:51.3 & 18.1 (r) & 7200  & 1&No\\
J1013+5615 & 10:13:36.4 & +56:15:36.4 & 18.5 (r) & 3600  & 5&Yes\\
J1049$-$0110 & 10:49:15.4 & $-$01:10:38.1 & 17.8 (r) & 4800  & 13&No\\
J1056+1208 & 10:56:48.7 & +12:08:26.8 & 17.9 (r) & 21300  & 18&Yes\\
J1155+0530 & 11:55:38.6 & +05:30:50.6 & 18.1 (r) & 7200  & 10&No\\
J1159+0112 & 11:59:44.8 & +01:12:07.0 & 17.3 (r) & 25000  & 22&Yes\\
J1200+4015 & 12:00:39.8 & +40:15:56.0 & 18.3 (r) & 10800  & 13&Yes\\
J1249$-$0233 & 12:49:24.9 & $-$02:33:39.7 & 17.7 (r) & 7300  & 10&Yes\\
J1310+5424 & 13:10:40.2 & +54:24:49.6 & 18.5 (r) & 10800  &11&Yes\\
J1313+1441 & 13:13:41.9 & +14:41:40.5 & 18.2 (r) & 10200  &9&Yes\\
J1417+4132 & 14:17:19.2 & +41:32:37.0 & 18.4 (r) & 25200 &  30&No\\
J1524+1030 & 15:24:30.1 & +10:30:32.0 & 18.1 (r) & 9000 & 6&Yes\\
J1552+4910 & 15:52:33.9 & +49:10:08.3 & 18.0 (r) & 9000 & 20&Yes\\
J1555+4800 & 15:55:56.9 & +48:00:15.0 & 19.1 (r) & 21600 & 6&Yes\\
J1604+3951 & 16:04:14.0 & +39:51:21.9 & 18.1 (r) & 10300 & 15&Yes\\
J1610+4724 & 16:10:09.4 & +47:24:44.5 & 18.8 (r) & 10800  &8&Yes\\
Q1755+578 & 17:56:03.6 & +57:48:48.0 & 18.3 (R) & 30400 &11&Yes\\
J2100$-$0641 & 21:00:25.0 & $-$06:41:45.0 & 18.1 (r) & 28000  &28&Yes\\
J2222$-$0945 & 22:22:56.1 & $-$09:46:36.0 & 18.0 (r) & 10800 &16&No\\
Q2230+02 & 22:32:35.3 & +02:47:55.1 & 18.0 (R) & 27600  &7&No\\
J2241+1225 & 22:41:45.1 & +12:25:57.1 & 17.9 (r) & 7200 & 6&Yes\\
J2340$-$0053 & 23:40:23.7 & $-$00:53:27.0 & 17.5 (r) & 15000 & 23&No \\
Q2342+34 & 23:44:51.1 & +34:33:46.8 & 19.1 (V) & 7200 & 11&No\\
\hline
\end{tabular}
\end{center}
\textsc{$^{a}$} The S/N is quoted near the observed position of the BII $\lambda 1362$ line.
\end{table*}

\begin{landscape}
\begin{table}
\caption{Boron, Oxygen, and Sulphur Column Densities and Abundances}
\label{tab:BS}
\begin{tabular}{lccccccccccc}

\hline
QSO & $z_{em}$&$z_{abs}$&logN(HI) &logN(BII)& logN(SII)& logN(OI) &[B/H] &[S/H] &[O/H] & [M/H] & M\\
\hline
J0008$-$0958 & 1.95 & 1.76753 & $20.85\pm0.15$ (1)& $<12.07$ & $15.83\pm0.02$ & $<18.03$& $<0.43$ & $-0.17\pm0.15$ & $<0.49$ & $-0.17\pm0.15$ & S\\
J0058+0115 & 2.49 & 2.00953 & $21.10\pm0.15$ (1)& $11.95\pm0.04$ & $15.41\pm0.01$ &\nodata& $0.06\pm0.16$ & $-0.84\pm0.15$ & \nodata & $-0.84\pm0.15$ & S\\
Q0201+36 & 2.49 & 2.46280 & $20.38\pm0.15$ (2)& \nodata & \nodata & \nodata & \nodata & \nodata & \nodata & $-0.35\pm0.15$ & Si\\
Q0458$-$02 & 2.29 & 2.03950 & $21.65\pm0.09$ (3)& $<12.53$ & \nodata & $<18.452$ &$<0.09$ &\nodata & $<0.18$ & $-1.10\pm0.10$ & Zn\\
FJ0812+3208 & 2.71 & 2.62593 & $21.35\pm0.15$ (4)& $11.43\pm0.08$ & $15.48\pm0.02$ & $<17.69$ & $-0.71\pm0.17$ & $-1.02\pm0.15$ & $<-0.35$ & $-1.02\pm0.15$ & S\\
J0927+1543 & 1.80 & 1.73113 & \nodata & $<12.96$& $<15.86$ & $<18.97$& \nodata & \nodata & \nodata & \nodata\\
J0927+5823 & 1.91 & 1.63515 & $20.40\pm0.25$ (2)& \nodata & $15.79\pm0.14$ & $<18.33$ &\nodata & $0.24\pm0.28$ & $<1.24$ &  $0.24\pm0.28$ & S\\
J1010+0003 & 1.40 & 1.26514 & \nodata & \nodata & \nodata & \nodata & \nodata &\nodata&\nodata & \nodata\\
J1013+5615 & 3.61 & 2.28400 & \nodata & \nodata & \nodata & \nodata & \nodata & \nodata & \nodata & \nodata\\
J1049$-$0110 & 2.12 & 1.65760 & $20.35\pm0.15$ (1) & \nodata & $15.47\pm0.01$ & $<18.11$ &\nodata & $-0.03\pm0.15$ & $<1.07$ & $-0.03\pm0.15$ & S\\
J1056+1208 & 1.92 & 1.60954 & $21.45\pm0.15$ (2) & $<12.21$ & $>15.66$ & $<18.20$ & $<-0.03$ & $>-0.94$ & $<0.06$ & $-0.32\pm0.16$ & Zn\\
J1155+0530 & 3.48 & 3.32607 & $21.05\pm0.10$ (4) & $<12.02$ & $15.35\pm0.003$ & $<17.66$ & $<0.18$ & $-0.85\pm0.10$ & $<-0.08$ & $-0.85\pm0.10$ & S\\
J1159+0112 & 2.00 & 1.94375 & $21.80\pm0.10$  (5)& $<12.04$ & $>15.16$ & \nodata & $<-0.55$ & $>-1.79$ & \nodata & $-1.30\pm0.11$ & Zn\\
J1200+4015 & 3.36 & 3.22000 & $20.65\pm0.15$ (4) & $<11.82$ & $15.37\pm0.01$ & $<17.79$ & $<0.38$ & $-0.43\pm0.15$ & $<0.45$ & $-0.43\pm0.15$ & S\\
J1249$-$0233 & 2.12 & 1.78085 & $21.45\pm0.15$ (6) & $<12.34$ & $15.50\pm0.02$ & $<18.23$& $<0.10$ & $-1.10\pm0.16$ & $<0.09$ & $-1.10\pm0.16$ & S\\
J1310+5424 & 1.93 & 1.80070 & $21.45\pm0.15$ (2) & $<12.39$ & $>15.96$ & $<18.39$ & $<0.15$ & $>-0.64$ & $<0.25$ & $-0.51\pm0.16$ & Zn\\
J1313+1441 & 1.88 & 1.79480 & \nodata  & $<12.21$ & $15.71\pm0.01$ & $<18.19$ &\nodata &\nodata &\nodata &\nodata \\
J1417+4132 & 2.02 & 1.95090 & $21.45\pm0.25$ (6) & $<12.02$ & $>15.8$ & $17.98\pm0.07$ & $<-0.22$ & $>-0.80$ & $-0.16\pm0.26$ & $-0.54\pm0.25$ & Zn\\
J1524+1030 & 2.06 & 1.94094 & \nodata & $<12.13$ & $>15.53$ & $<18.05$&\nodata &\nodata &\nodata & \nodata & \\
J1552+4910 & 2.04 & 1.95987 & \nodata & $<11.95$ & $15.34\pm0.004$ & $<17.96$&\nodata &\nodata &\nodata &\nodata & \\
J1555+4800 & 3.30 & 2.39089 & $21.50\pm0.15$ (4) & \nodata & $>15.88$ &\nodata & \nodata& $>-0.77$&\nodata & $-1.22\pm0.20$ & Fe\\
J1604+3951 & 3.13 & 3.16400 & $21.75\pm0.20$ (4) & $<11.71$ & $15.70\pm0.04$ & $<17.69$ & $<-0.83$ & $-1.20\pm0.20$ & $<-0.75$ & $-1.20\pm0.20$ & S\\
J1610+4724 & 3.22 & 2.50661 & $21.00\pm0.15$ (4) & $<12.38$ & $>16.01$ & $<18.35$ & $<0.59$ & $>-0.14$ & $<0.66$ & $-0.08\pm0.16$ & Zn\\
Q1755+578 & 2.11 & 1.97110 & $21.40\pm0.15$ (7) & $<11.91$ & $>15.97$ & $<17.88$ & $<-0.28$ & $>-0.57$& $<-0.21$ & $-1.04\pm0.15$ & Fe\\
J2100$-$0641 & 3.14 & 3.09130 & $21.05\pm0.15$ (4) & $<11.78$ & $15.64\pm0.002$ & $<17.71$ & $<-0.06$ & $-0.56\pm0.15$ & $<-0.03$ & $-0.56\pm0.15$ & S\\
J2222$-$0945 & 2.93 & 2.35430 & $20.55\pm0.15$ (4) & $<12.05$ & $15.33\pm0.02$ & \nodata& $<0.71$ & $-0.37\pm0.15$ & \nodata & $-0.37\pm0.15$ & S\\
Q2230+02 & 2.15 & 1.86440 & $20.85\pm0.08$ (3) & $<12.56$ & \nodata & $<18.55$  &$<0.92$&\nodata& $<1.01$ & $-0.66\pm0.09$ & Zn\\
J2241+1225 & 2.63 & 2.41800 & $21.15\pm0.15$ (4) & $<12.01$ & $14.94\pm0.04$ & $<17.98$& $<0.07$ & $-1.36\pm0.16$ & $<0.14$ & $-1.36\pm0.16$ & S\\
J2340$-$0053 & 2.09 & 2.05452 & $20.35\pm0.15$ (4) & $<11.60$ & $14.95\pm0.004$ & $<17.56$& $<0.46$ & $-0.55\pm0.15$ & $<0.52$ & $-0.55\pm0.15$ & S\\
Q2342+34 & 2.92 & 2.90899 & $21.10\pm0.10$ (8) & $<11.96$ & $15.17\pm0.01$ & $<17.96$& $<0.07$ & $-1.08\pm0.10$ & $<0.17$ & $-1.08\pm0.10$ & S\\
\hline
\end{tabular}

\end{table}
\textsc{HI References} --
		(1) \cite{HerbertFort06}; 
		(2) \cite{Kaplan10}; 
		(3) \cite{Pettini94}; 
		(4) \cite{Prochaska09}; 
		(5) \cite{Kanekar09}; 
		(6) This paper; 
		(7) \cite{Jorgenson06}; 
		(8) \cite{Prochaska03H}; 
\end{landscape}

\section{Abundance determination}
\label{sec:abund}

\begin{table}
\begin{center}
\caption{Wavelengths and Oscillator Strengths of Transitions}
\label{tab:fvals}
\begin{tabular}{lcc}
\hline
Element &Wavelength (\AA)& $f$\\
\hline
SII &1250.584 &5.453$\cdot10^{-3}$\\
SII &1253.811 &1.088$\cdot10^{-2}$\\
SII &1259.519 &1.624$\cdot10^{-2}$\\
OI &1355.598 &1.248$\cdot10^{-6}$\\
BII &1362.461 &9.870$\cdot10^{-1}$\\
\hline
\end{tabular}
\end{center}
\textsc{Reference} -- \cite{Morton03}
\end{table}

\begin{table}
\begin{center}
\caption{Solar abundances}
\label{tab:solar}
\begin{tabular}{lcc}
\hline
Element & logN(X/H)$_\odot$+12 & Source\\
\hline
B & $2.79\pm0.04^{a}$ & Meteoritic\\
C & $8.43\pm0.05$ & Photospheric\\
N & $7.83\pm0.05$ & Photospheric\\
O & $8.69\pm0.05$ & Photospheric\\
S & $7.15\pm0.02$ & Meteoritic\\
\hline
\end{tabular}
\end{center}
\textsc{Reference} -- \cite{Asplund09}\\
$^{a}$ The measured solar abundance of boron
by \cite{Asplund09} is higher by $\sim0.3$ dex relative to the
local ISM and B-type stars that have been used in previous works such
as \cite{Venn02}, see also the discussion in  \cite{Cunha97}.  The
impact of the uncertainty in the solar boron abundance is discussed
in Section \ref{sec:Discussion}.
\end{table}

All metal column densities were obtained using the apparent optical depth
method (AODM) outlined by \cite{Savage91}. The limits for the optical depth
integrations were chosen to contain the absorption profile common for several non-contaminated lines. These limits are shown by the vertical dotted lines in Figures \ref{fig:ohohspec}, \ref{fig:oheightspec}, and \ref{fig:onefourspec}
(although these have been adjusted from their fiducial values in
cases of suspected contamination, which we discuss on a case by
case basis below). The errors quoted in Table \ref{tab:BS} were determined from the photon noise. Continuum errors are typically neglected for strong transition lines as the error is dominated by photon noise. However for the weak detections presented in Section \ref{subsec:detections}; continuum fitting errors may play a role in the abundance determination and are further discussed on a case by case basis in Section \ref{subsec:continuum}.

For the majority of the absorbers in our sample, HI column
densities have been previously determined either directly from the
SDSS spectra \citep[10 DLAs,][]{Prochaska09}, follow-up spectroscopy obtained
with either Keck II/ESI  \citep[3 DLAs,][]{HerbertFort06} or MMT/BCS \citep[4 DLAs,][]{Kaplan10}, or were adopted from previous studies (5 DLAs). For the remaining two DLAs, we have determined the N(HI) directly
from the HIRES spectrum.  HI column densities were determined by fitting a fully damped Voigt
profile to the Ly$\alpha$ transition using a function with the \textsc{XIDL} software. Six DLAs have no measured N(HI) currently
available; these are lower redshift DLAs for which neither the
original SDSS spectrum nor the new HIRES data cover the Ly$\alpha$ transition.
The systems without an N(HI) measurement are still included in our
sample, since a measurement of the B/S or B/O can still help us
to constrain production mechanisms.

To estimate $3\sigma$ upper limits in the case of non-detections, we
took the value of the full width at half maximum (FWHM) of the most
prominent feature in the absorption profile of a strong
transition. This feature should correspond to the most easily detected
absorption of any element in the DLA. From the S/N ratio
of the spectrum at the location of the absorption line (see Table \ref{tab:MSDLAs}\footnote{The blank S/N ratio entry represents no coverage in the spectrum and is included for completeness. DLAs classified as MSDLAs (based on the silicon and zinc abundances from Berg et al., in prep.) are flagged in the final column.}), the rest frame
equivalent width (W; at $n\sigma$ significance) of a line at redshift $z$ is calculated from 

\begin{equation}
W=\frac{n\cdot FWHM}{S/N\cdot(1+z)}.
\label{eq:eqw}
\end{equation}

From equation \ref{eq:eqw}, the column density of a species
with rest wavelength $\lambda_{0}$ and oscillator strength $f$ is
calculated from

\begin{equation}
N = \frac{\pi e^{2}}{m_{e}c^{2}}\frac{W}{f\lambda_{0}^{2}}.
\label{eq:N_lim}
\end{equation} 

 All wavelengths, oscillator strengths \citep{Morton03}, and solar values \citep{Asplund09} adopted are listed in Tables \ref{tab:fvals}  and \ref{tab:solar}, respectively. The measured column
densities and derived upper limits are presented in Table
\ref{tab:BS}, along with the quasar emission redshift ($z_{em}$),
DLA absorption redshift ($z_{abs}$), and HI column densities of the DLAs. All $3\sigma$ upper limits presented are only obtained for the spectra where little or no contamination is present using the S/N near the absorption. Otherwise, no upper limit was obtained despite S/N being reported for all DLAs near the line in Table \ref{tab:MSDLAs}. 

In this study, we focus solely on the abundances of B, O, and S.  We
assume that the gas phase abundances that we have measured are
representative of the total elemental abundance in the DLA.  Oxygen
and sulphur are indeed non-refractory and little depleted onto dust
grains \citep{Savage96}.  Although boron is
depleted in dense Galactic disk clouds \citep{Howk00}, sight-lines
through low density gas should yield a robust estimate of the boron
abundance \citep{Ritchey11}.  The abundance pattern of DLAs
\citep{Pettini00}, typically low molecular
content \citep{Ledoux03} and high spin
temperature \citep{Ellison12} indicate that most DLA 
sight-lines do not intersect cold, high density clouds. Hence
the boron abundances measured in these systems should not be greatly
affected by depletion. However, due to the generally higher metallicities of the MSDLAs in our sample, we may be probing higher density sight-lines which may have enhanced dust depletion. We also assume that the measured ionization
state for each element is the dominant one and that no ionization
corrections are necessary.  We discuss explicitly the effect of
possible depletion and ionization effects on our conclusions in Section
\ref{sec:Discussion}.

\subsection{Possible Detections}
\begin{figure}
\begin{center}
\includegraphics[width=0.5\textwidth]{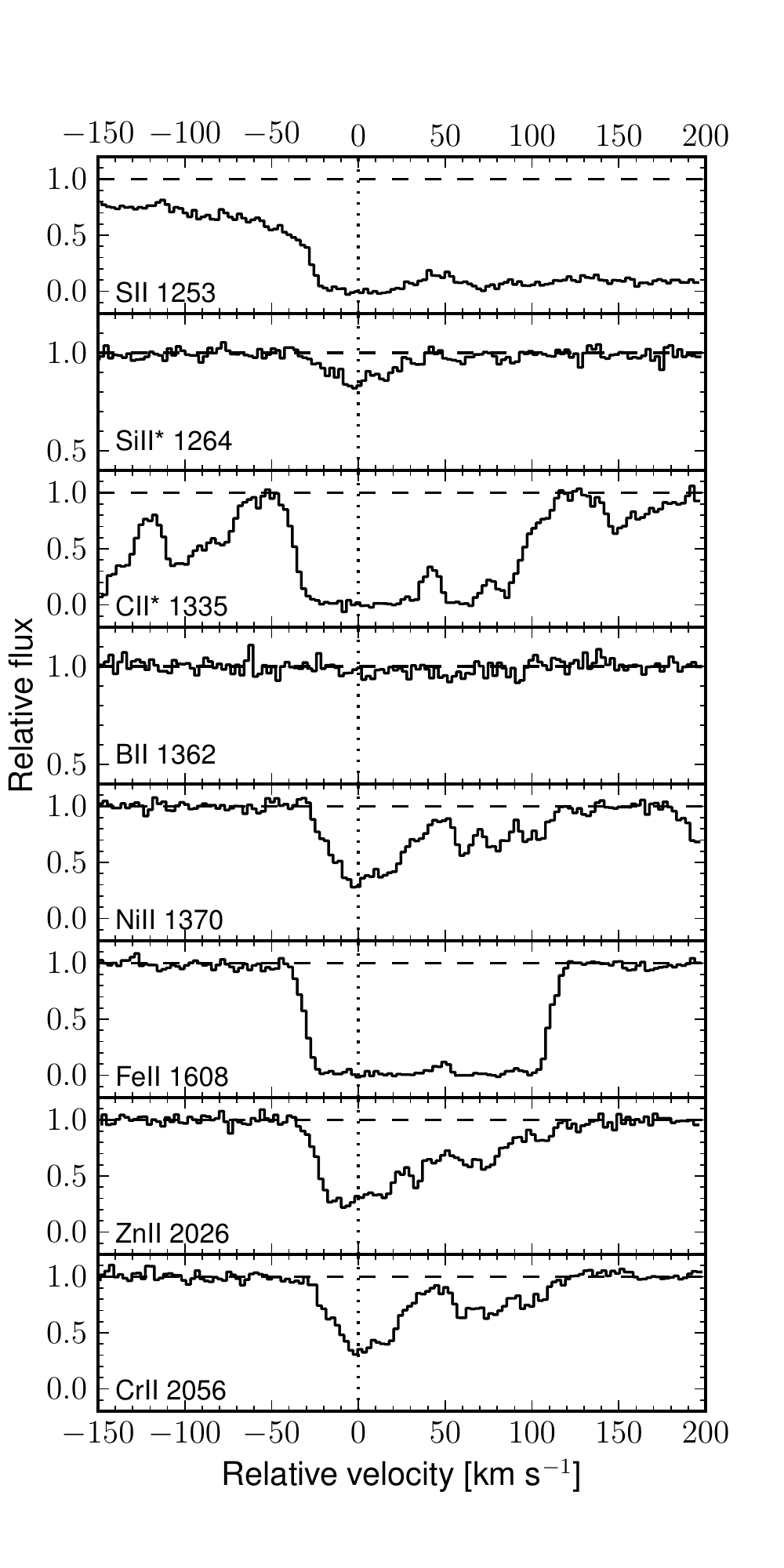}
\caption{Absorption profiles for various species in J1417+4132. Identical absorption profiles are seen for all singly ionized species (other than SII; which is lost by broad absorption). Again, we note that there seems to be absorption at the Si II$^{\star}$ $\lambda\lambda$ 1264.}
\label{fig:onefouradd}
\end{center}
\end{figure}

Three of the DLAs in our sample have visibly significant absorption features
at the expected wavelength of BII $\lambda$ 1362. The profiles from
each are discussed below, including possible sources of blending and
contamination. Figures \ref{fig:ohohadd}, \ref{fig:oheightadd}, and \ref{fig:onefouradd} show the absorption profiles for several different species; whereas Figures \ref{fig:ohohspec}, \ref{fig:oheightspec}, and \ref{fig:onefourspec}
zoom in and show the profiles for both boron and oxygen. For comparison
the scaled profiles of moderately strong lines are shown for visual
guidance of the general shape of absorption profiles for these DLAs\footnote{The scaling is done by matching the area integrated within the AODM bounds for the reference metal line profile to the area determined for boron or oxygen.}.

 \label{subsec:detections}
\subsubsection{J0058+0115}
\begin{figure}
\begin{center}
\includegraphics[width=0.5\textwidth]{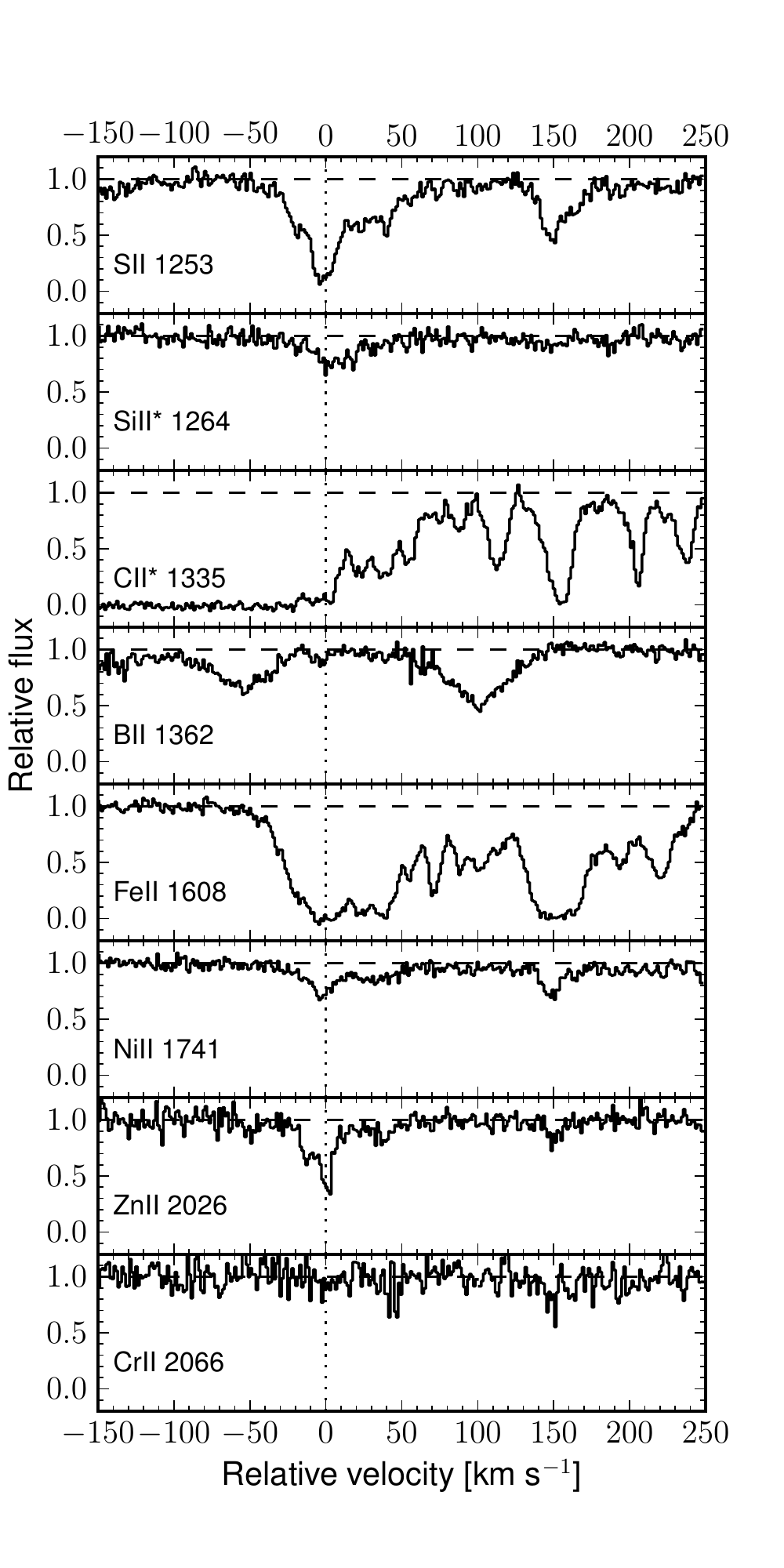}
\caption{Absorption profiles for various singly ionized species in J0058+0115. From panel to panel, it is apparent that the absorption profiles for all the species (other than CII$^{\star}$) traces out an identical shape for various absorption strengths.}
\label{fig:ohohadd}
\end{center}
\end{figure}

\begin{figure}
\begin{center}
\includegraphics[width=0.5\textwidth]{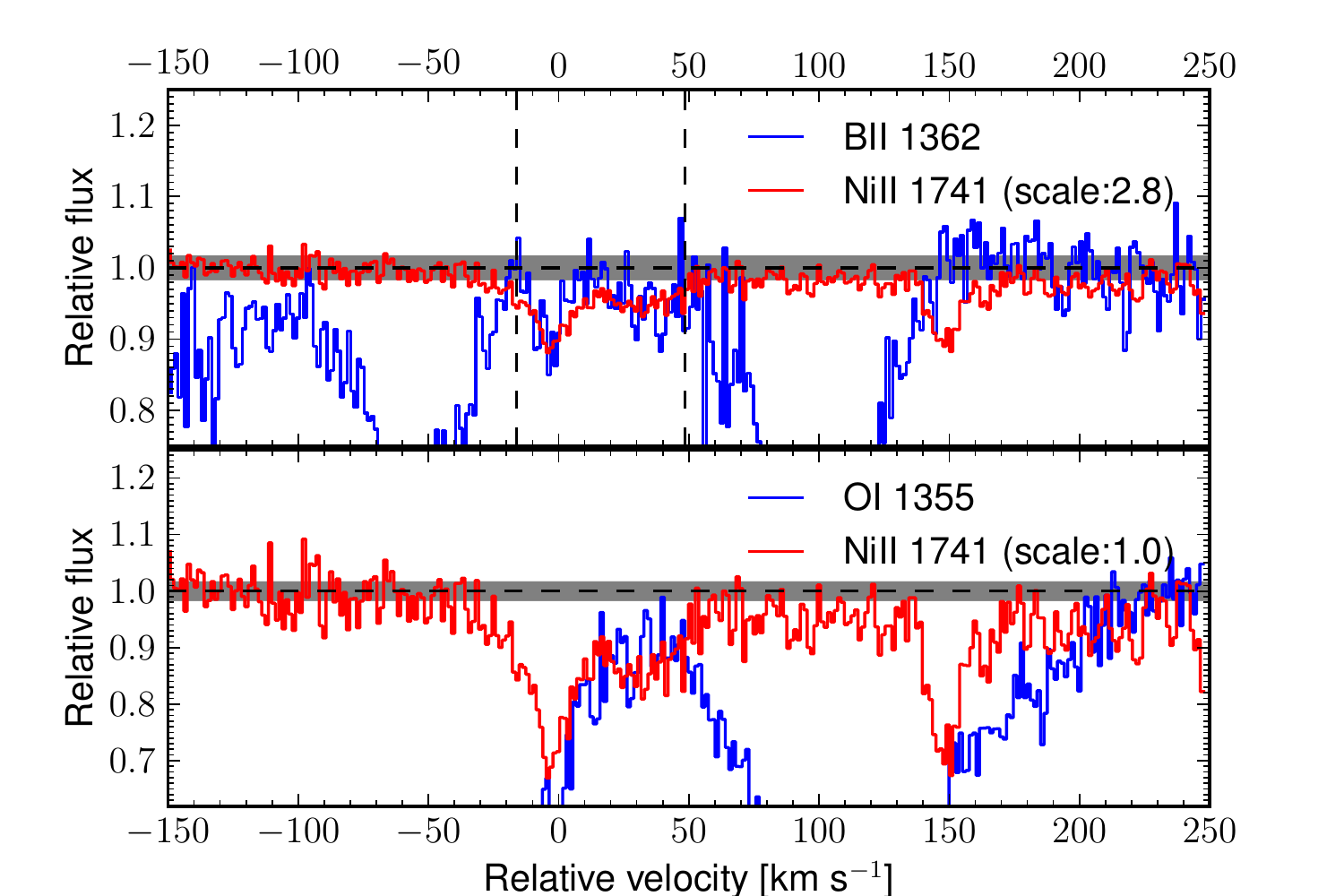}
\caption[J0058+0115]{The absorption profile of BII $\lambda1362$
(top, blue), OI $\lambda1355$ (bottom, blue) and SII $\lambda1250$
(red) for J0058+0115. The black horizontal dashed line indicates
the continuum, and the black vertical dotted lines are the bounds for
the apparent optical depth integration for the species in blue. The
nickel profile is scaled down to match the intensity of the boron and
oxygen lines by matching the areas within the AODM bounds of the absorption. Oxygen is not detected in this DLA as it is blended within the Ly$\alpha$ forest. Overall, the scaled profile matches that of boron absorption. The grey regions mark where the continuum has been over and underestimated for determining errors in the continuum (see Section \ref{subsec:continuum}).}
\label{fig:ohohspec}
\end{center}
\end{figure}

Figure \ref{fig:ohohadd} displays the absorption features for various transition lines in J0058+0115, whereas Figure \ref{fig:ohohspec} shows an absorption feature at the expected
wavelength of BII $\lambda$ 1362 and OI $\lambda$ 1355.  The scaled nickel line is overplotted in Figure \ref{fig:ohohspec} for comparison. The shape of the boron absorption is in very good agreement with that
of the Ni II $\lambda$ 1741 line, hence we report a possible detection of boron for the first time in this DLA, where the measured equivalent width is $3.55\sigma$ significant\footnote{The significance is calculated by combining Equations \ref{eq:eqw} and \ref{eq:N_lim}, then solving for the number of sigma (n) needed to obtain the derived AODM abundance. The FHWM was measured from a prominent absorption component of a strong metal line in the spectrum, and is equivalent to the value used in the calculation for the upper limits.}. However, it is not feasible to rule out the possibility that this feature is actually a weak Ly$\alpha$ line.
The OI $\lambda$ 1355 line is completely blended with
the Ly$\alpha$ forest and we therefore do not report a limit.

\subsubsection{FJ0812+3208}
\label{sec:oheight}
\begin{figure}
\begin{center}
\includegraphics[width=0.5\textwidth]{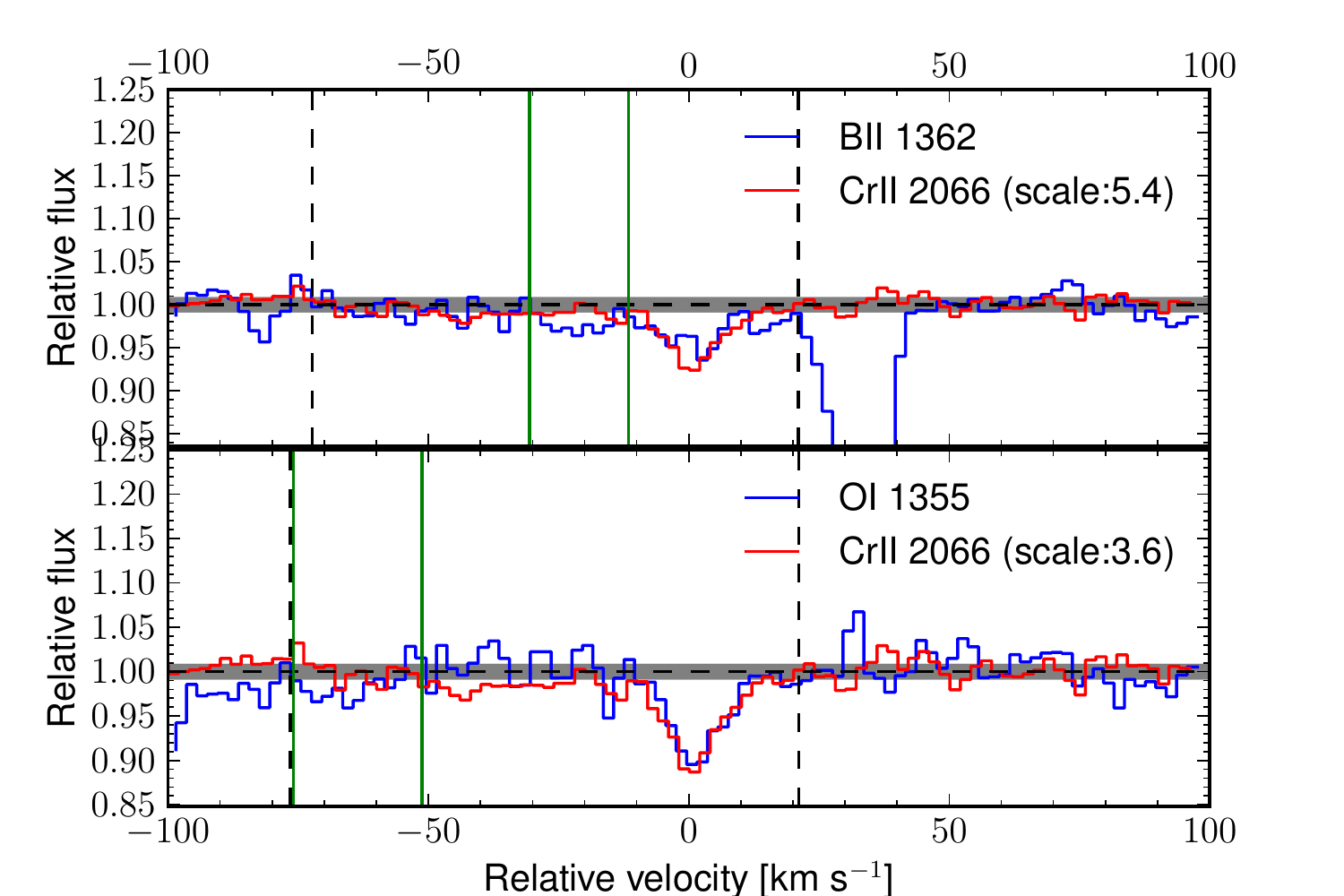}
\caption[FJ0812+3208]{The absorption profile of boron and oxygen for 
FJ0812+32 following the notation in Figure \ref{fig:ohohspec}.
 Extra components that are not present in the chromium profile are bounded 
by the solid green vertical lines and are not included in the abundance determination. Boron appears to be blended by a feature at 
$v\sim-20$ \kms{}, which we have removed from the abundance determination. The oxygen absorption at $\sim-40$ \kms{} does not appear, despite the strength of the component at $\sim0$ \kms{} and high S/N. See Section
\ref{sec:oheight} for a discussion of possible contamination of the oxygen line.The grey regions mark where the continuum has been over and underestimated for determining errors in the continuum (see Section \ref{subsec:continuum}).}
\label{fig:oheightspec}
\end{center}
\end{figure}

Boron was detected in the FJ0812+3208 sight-line at a
redshift of $z=2.626$ by \cite{Prochaska03} with a column density
logN(B)$=11.43$. The data presented here for this sight-line have
been improved from that presented in \cite{Prochaska03} by
the addition of $\sim10$ hours of HIRES data \citep{Jorgenson09}.
Both the BII $\lambda$ 1362 and OI $\lambda$ 1355 lines are shown in Figure
\ref{fig:oheightspec}, with scaled CrII $\lambda$ 2066 shown for comparison. It can be seen that the CrII line has two components: the stronger one centred at $v\sim0$ \kms{} and a weaker component at $\sim-40$ \kms{}. The CrII profile is used to define the velocity range over which we integrate the optical depth in both OI and BII (vertical dotted lines). The BII $\lambda$1362 line is well aligned with CrII $\lambda$ 2066. We determine a column density log(B/H)=$11.43\pm0.08$ (measured at $3.84\sigma$ significance) which is in agreement with the result of \cite{Prochaska03}.   
Figure \ref{fig:oheightspec} shows that there is an additional
BII $\lambda$1362 component at $v\sim-20$ \kms{} (bounded by the solid
vertical green lines) that is not seen in CrII. 
We have excluded it in the abundance determination. The additional absorption at $v\sim +40$ \kms{} is unrelated absorption and has been excluded from the analysis.
The oxygen profile is detected only in the strongest
component at $v\sim0$ \kms{}, but not the weaker component seen in other lines at
$v\sim-40$ \kms{} (see Figure \ref{fig:ohohadd}).  Given the S/N, we might have expected a detection (albeit weak) of the $v\sim-40$ \kms{} component, if the column densities of the two components have the same relative ratio in oxygen as sulphur. We must therefore consider the possibility that the oxygen detection may be contaminated by a different species. Indeed,
if the absorption was entirely due to OI, it would yield a [S/O] = $-0.67$, which
is a surprising discrepancy from the solar value.  However, we note
that in \cite{Prochaska03} [S/O] is determined to be [S/O] = $-0.33$
which is closer to the relative abundance in nearby disk stars \citep[e.g.][see Section \ref{sec:SO}]{Reddy03}.  
This discrepancy with \cite{Prochaska03} is due to both a lower N(SII) 
and a higher N(OI) in our study. To be conservative and because we have a 
robust detection of sulphur for our analysis; we report oxygen as an upper 
limit from the equivalent width of the detected absorption feature in recognition of possible contamination.

\begin{figure}
\begin{center}
\includegraphics[width=0.5\textwidth]{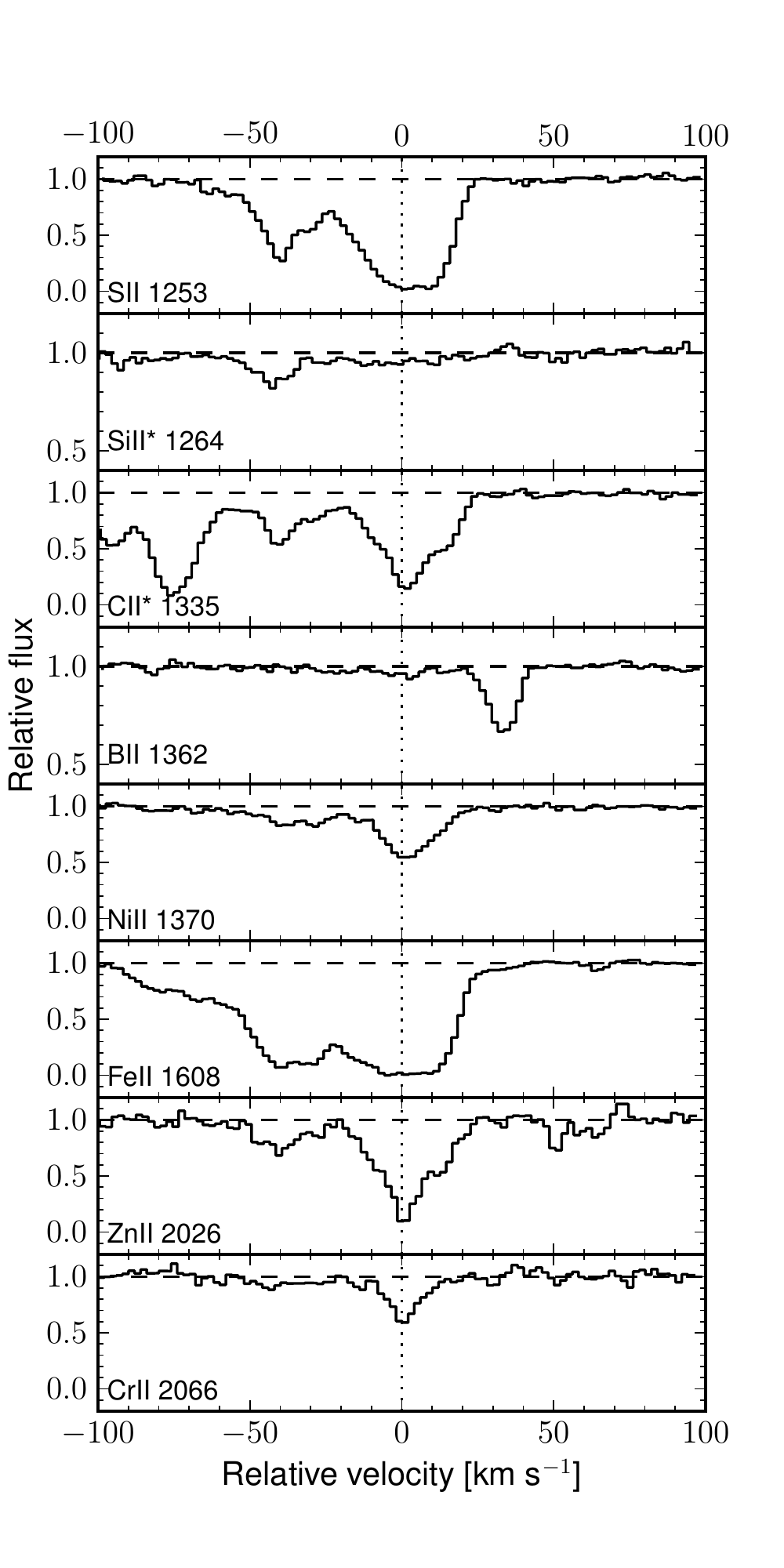}
\caption{Similar to Figure \ref{fig:ohohadd}, the same absorption profile is seen for all singly ionized species detected in FJ0812+3208.}
\label{fig:oheightadd}
\end{center}
\end{figure}

\subsubsection{J1417+4132}
\begin{figure}
\begin{center}
\includegraphics[width=0.5\textwidth]{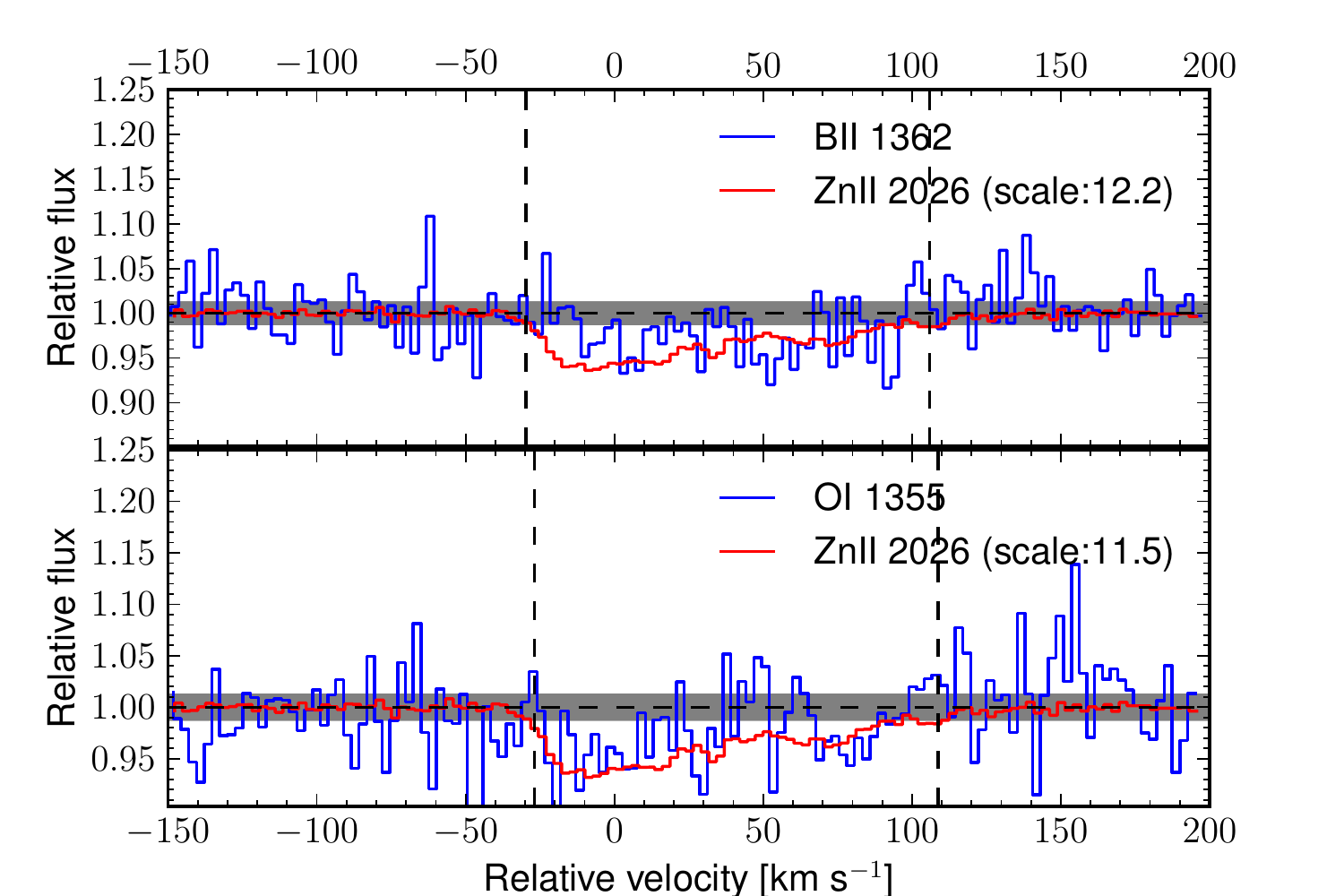}
\caption[J1417+4132]{The absorption profiles of boron, oxygen and zinc
for J1417+4132 following the same notation as in Figure
\ref{fig:oheightspec}.   The grey regions mark where the continuum has been over and underestimated for determining errors in the continuum (see Section \ref{subsec:continuum}). The absorption is only significant to $2.97\sigma$, so it is treated as an upper limit.}
\label{fig:onefourspec}
\end{center}
\end{figure}

J1417+4132 contains a possible detection of boron (see Figure
\ref{fig:onefourspec}) that we report here for the first time. The scaled ZnII $\lambda$2026 line is shown for comparison.
The detection is challenging in this case because of the broad,
shallow nature of the profile; but there is a clear
drop below the continuum
over the same velocity range where the boron profile would lie for
both OI and BII. However, the feature is broad and shallow, with a total equivalent width significant to $2.97\sigma$, making it a very tentative detection. Despite being nearly $3\sigma$, we adopt the column density as an upper limit due to the challenging nature of the detection. However, a discussion on how the continuum errors may affect the yields of both boron and oxygen is furthered in Section \ref{subsec:continuum} for completeness. Unfortunately, all of the SII lines are saturated or severely blended, so we can only determine a lower limit for N(SII).

\subsection{Continuum errors}
\label{subsec:continuum}

Due to the weakness of the putative boron features, continuum errors may play a role for the three DLAs discussed above. To check how much the continuum placement may affect our results, we artificially placed the continuum at higher and lower values to estimate the difference in column density resulting from bad continuum fitting. This is represented by the top and bottom edges of the grayed regions in Figures \ref{fig:ohohspec}, \ref{fig:oheightspec}, and \ref{fig:onefourspec}. These offsets correspond to 40\% of the inverse of the S/N of the spectrum near the absorption. The resulting differences in the column densities ($\Delta$logN) for the associated continuum offsets are shown in Table \ref{tab:cont} for both boron and oxygen. In general, the change in column density for both oxygen and boron ($<0.18$ dex) is about the same magnitude or smaller than the typical error in the hydrogen column density (0.15--0.25 dex; see Table \ref{tab:BS}). Specifically, for J1417+4132 (where $\Delta$logN(B) is $\sim0.17$ dex) the difference is small relative to the overall error budget in [B/H] ($0.26$ dex) which is primarily dominated by the uncertainty in the hydrogen column ($0.25$ dex). However the change in column density for boron in both J0058+0115 and FJ0812+3208 is of the same order as the hydrogen column error.  The uncertainty in the total error budgets in [O/H] and [B/H] ($\sigma_{O}$ and $\sigma_{B}$; respectively) have been recalculated from Table \ref{tab:BS} to include this continuum error and are shown in Table \ref{tab:cont}. Including the continuum fitting error in the total error budget results in at most a 0.07 dex change, which is relatively small. Based on these results, we conclude that continuum errors do not have a significant contribution in any of these three DLAs.

\begin{table*}
\begin{center}
\caption{Continuum errors}
\label{tab:cont}
\begin{tabular}{lccccccc}
\hline
QSO & Continuum offset & $\Delta$logN(B)$^{a}$ & $\Delta$logN(O)$^{a}$ & $\sigma_{B_{orig}}$ $^{b}$ & $\sigma_{O_{orig}}$ $^{b}$ & $\sigma_{B_{new}}$ $^{c}$ & $\sigma_{O_{new}}$ $^{c}$ \\
\hline
J0058+0115 & $\pm0.013$ & $\pm0.14$ & \nodata & $\pm0.16$ & \nodata & $\pm0.21$ & \nodata \\
FJ0812+3208 & $\pm0.007$ & $\pm0.18$ & \nodata & $\pm0.17$ & \nodata & $\pm0.24$ & \nodata \\
J1417+4132 & $\pm0.014$ & ($\pm0.17$)$^{d}$ & $\pm0.15$ & ($\pm0.26$) & $\pm0.26$ & ($\pm0.31$) & $\pm0.30$ \\
\hline
\end{tabular}
\end{center}
\textsc{$^{a}$} The error in the column density resulting from offsetting the continuum.\\
\textsc{$^{b}$} The error budget in the abundance [X/H] from the AODM calculation (Table \ref{tab:BS}).\\
\textsc{$^{c}$} The total error budget in the abundance [X/H] including the error from continuum placement.\\
\textsc{$^{d}$} The errors in parentheses represent the derived AODM errors from assuming the detection is real.\\
\end{table*}

\section{Discussion }
\label{sec:Discussion}

\subsection{Sulphur-oxygen relation}
\label{sec:SO}

\begin{figure}
\begin{center}
\includegraphics[width=0.5\textwidth]{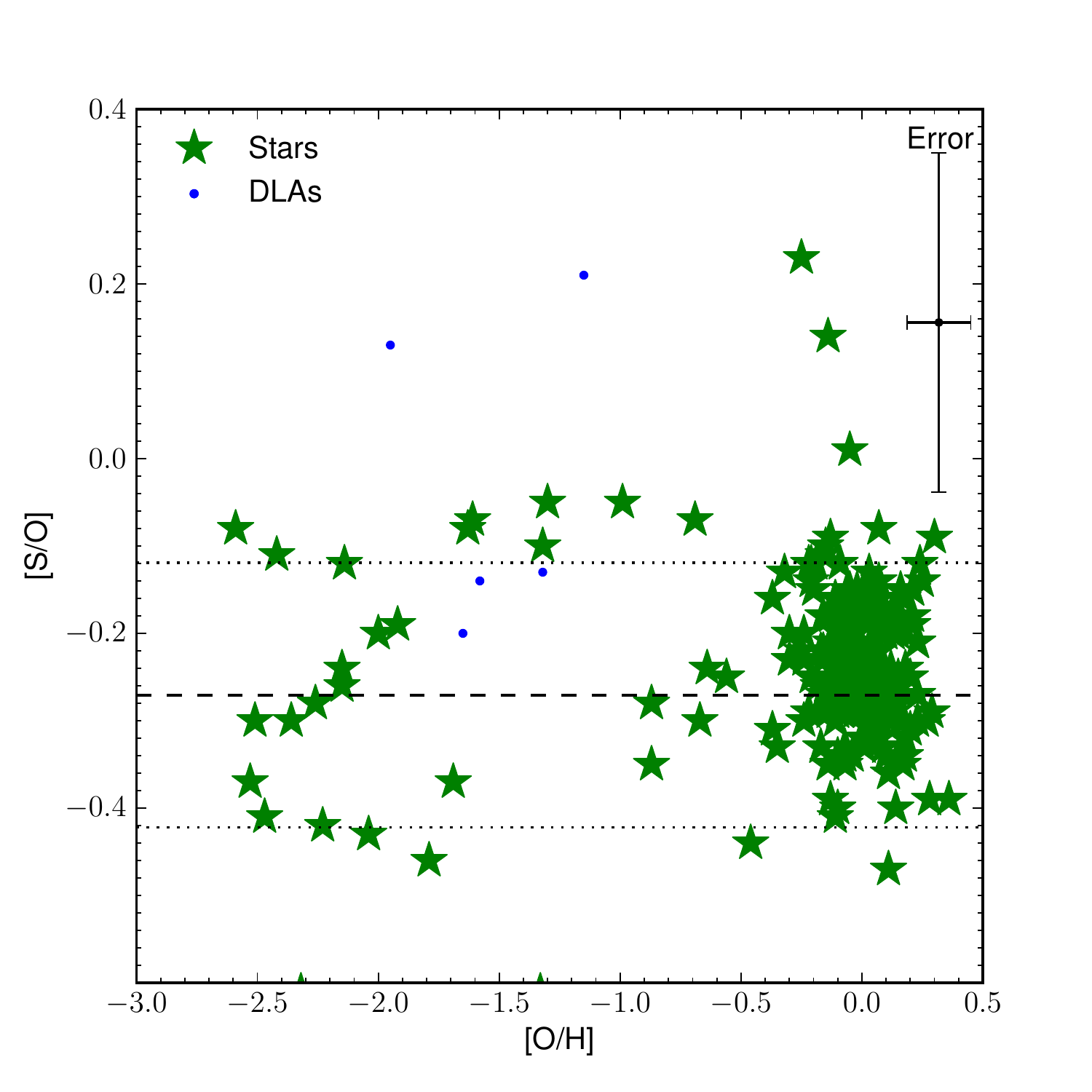}
\caption[Sulphur-Oxygen relation]{ [S/O] is plotted as a function of
[O/H] for stellar data to determine whether these two element track each other in stars. The data plotted includes the entire literature compilation presented in Table \ref{tab:stellarBSO}, including both stars and DLAs. The dashed line is the mean value of [S/O]$=-0.27\pm0.15$ (with
dotted lines as $1\sigma$ errors) of the plotted data. The error bar in the top right shows the typical uncertainty in the stellar abundances.}
\label{fig:SO}
\end{center}
\end{figure}

In order to overcome the difficulties of determining oxygen abundances in 
stars\footnote{Although the comparison with the ISM is more realistic; we include stars as well since we are motivated by determining the nucleosynthetic origin of boron. Previous studies of stars and the ISM provide this information, hence both are included.} and DLAs, we investigate in this work the possibility of using
sulphur as an alternative alpha-capture reference element. The
comparison of boron with sulphur in the DLAs and stellar data therefore 
hinges on whether the sulphur abundance intrinsically tracks oxygen in stars. Other than stars containing boron, the stellar literature sources were selected on the basis of what lines and corrections were used in attempts to homogenize the sample and remove systematic errors (see Table \ref{tab:stellarBSO}\footnote{Abundances have been
converted to the \cite{Asplund09} scale.}). For oxygen, we only selected literature samples that used [O I], UV OH lines with proper continuum placement, or the OI triplet with non-LTE corrections. Following the analysis done by \cite{Caffau05}, sulphur studies that used the S I multiplet 1 (with non-LTE corrections), 6, and 8 for abundance determinations were selected. DLAs not present in our sample which contained both oxygen and sulphur abundances are included as well solely for comparison. 
 
Figure \ref{fig:SO}
shows the relation between [S/O] and [O/H] for the literature sample used. DLAs are not included in determining the mean offset as they are only shown for interest. Although the scatter about the mean appears substantial, most of the stellar data are consistent within 1.5$\sigma$ of the mean.  Having a constant offset in stars suggests that the interchange of oxygen and sulphur is valid, and should not change the slope of the best fit line between B/O and B/S. Due to the variety in non-LTE corrections, model atmospheres, and choice of lines; a more homogeneous dataset of oxygen and sulphur in stars should tighten the relation.

\subsection{Boron-oxygen relation}
\label{sec:BO}

\begin{figure*}
\begin{center}
\includegraphics[width=\textwidth]{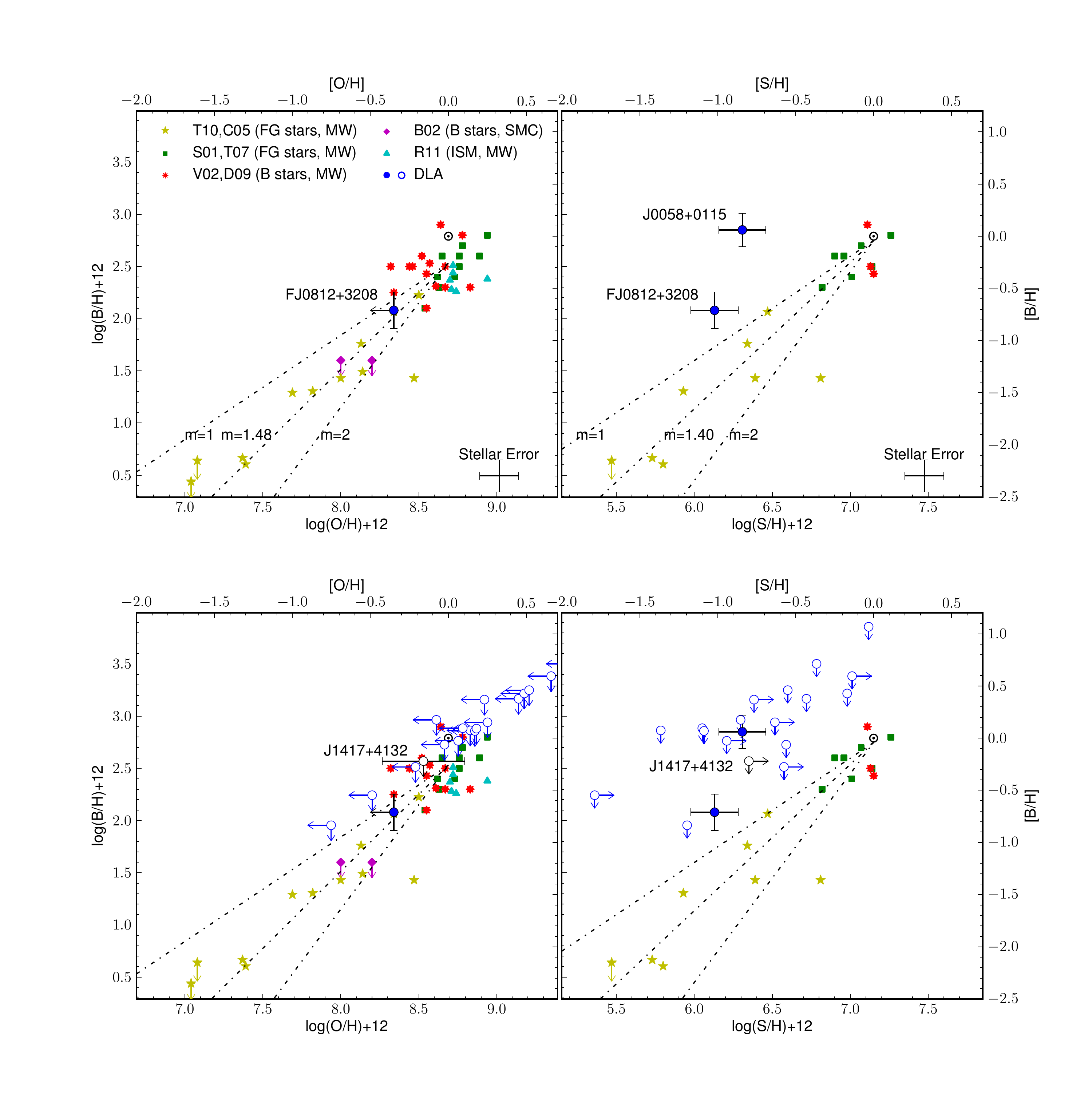}
\caption[Boron-Sulphur and Boron-Oxygen relations]{B/H is plotted as a
function of O/H (left panels) and S/H (right panels). The stellar boron data is from \cite{Smith01} (S01), \cite{Tan10} (T10), \cite{Venn02} (V02), \cite{Brooks02} (B02). The sulphur abundances for the Galactic data were adopted from \cite{Caffau05} (C05), \cite{Takeda07} (T07), and
\cite{Daflon09} (D09).  Open circles are used for any boron upper limits in DLAs, whereas the filled circles represent possible detections. The top set of panels only show the data of the two identified DLAs (J0058+0112 and FJ0812+3208), whereas the bottom panels include all upper limits as well. Typical errors in the stellar data are shown in the bottom right of the top panels. The solar \citep{Asplund09} scale is
included along the top and right axes in both panels. The best fit
equations are: $[B/H] = (1.48\pm0.059)[O/H] - (0.26\pm0.110)$ (left); 
$[B/H] = (1.40\pm0.117)[S/H] - (0.04\pm0.121)$ (right). As the \cite{Asplund09} values
are included in the fit, we do not force the lines to pass through the origin.}
\label{fig:BS}
\end{center}
\end{figure*}

In Figure \ref{fig:BS} we present our main scientific results, the
comparison of DLA and Galactic boron abundances
plotted as a function of oxygen (left) and sulphur (right). Both
panels contain the primary and secondary lines (of slopes $m=1$ and
$m=2$, respectively) and the best fit lines we have derived from
a linear least squares fit:

\begin{equation}
[B/H] = (1.48\pm0.059)[O/H] - (0.26\pm0.110)
\end{equation}

and

\begin{equation}
[B/H] = (1.40\pm0.117)[S/H] - (0.04\pm0.121).
\end{equation}

These best fit lines only include all detections in the Milky Way from \cite{Smith01} and 
\cite{Tan10}, as well as the
solar value \citep{Asplund09}\footnote{As the solar point is included, we do not force the primary, secondary, and best-fit lines through it.}. We also included in the fit
B stars that appeared not to be
affected by rotational mixing\footnote{Based on stars with log(N/H)$<-4.2$;
log(B/H)$>-10.0$ from \citep{Venn02}. Note that there still may be boron 
depletion \citep{Mendel06} as boron depletion occurs before nitrogen 
enrichment.} and ISM sight-lines that contained warm gas \citep{Ritchey11}\footnote{The \cite{Ritchey11} sample contains a compilation 56 sight-lines. Only the 6 lowest density sight-lines from the sample were chosen for our dataset as they are relatively unaffected by dust depletion.}. 
Both primary and secondary lines are forced to go through
the origin of the best fit. The B-O relation agrees with the best fit
equations presented in \cite{Smith01} and \cite{Ritchey11} (which have $m=1.39\pm0.08$,  $1.5\pm0.1$; respectively).

For the first time, we can now place high redshift data points on the
boron abundance plot.  The traditional presentation of
oxygen and boron abundances is shown in the left panels of  Figure \ref{fig:BS}.
Note that, for convenience, axes are labeled for both B/H and [B/H]
(similarly for oxygen).
Only the potential detections are plotted in the top panels of Figure \ref{fig:BS}, whereas all data (including the limits) are shown in the bottom panels. The DLA boron upper limits are above the primary ($m=1$)
line and therefore do not provide useful constraints on boron
production in the DLAs.  Moreover,
we have no further measurements of O/H, so the DLA points are
limits in both quantities.  Note that these
limits cluster parallel to the $m=1$ line because the BII $\lambda$
1362 and OI $\lambda$1355 lines are so close in wavelength that their
S/N ratios are essentially identical. In combination with the same assumed FWHM of the strongest feature (see Section \ref{sec:abund}) and similar $f\cdot(\frac{X}{H})_{\odot}$, Equation \ref{eq:N_lim} implies nearly equal abundances limits of boron and oxygen for each system.

Relative to oxygen, FJ0812+3208 is still consistent with the primary line. However, we have conservatively reported the oxygen abundance in FJ0812+3208 as an upper limit. 

Turning now to the B-S abundances which are shown in the right
panels of Figure \ref{fig:BS}.  In addition to circumventing the
problem of uncertain stellar O/H abundances, it can be seen
that sulphur has the added advantage of being measurable in many
of the DLAs.  Nonetheless, the boron non-detections all lie
above the primary line, so the production of boron is \emph{not constrained}.
However, both J0058+0115 and FJ0812+3208, which have robust detections in both B and S
are above the primary line, with the inclusion of the continuum errors mentioned in Section \ref{subsec:continuum}.

The data presented here reveal the \emph{possibility} of super-primary
boron abundances in two DLAs (indeed, the only two where boron is tentatively detected) at high $z$.  Although these two DLAs are unlikely to be representative of the DLA population,
their high N(HI) and relatively high metallicities are unusual
but make them ideal for a conceptual study of boron.  We now consider
possible corrections that are required to the abundances that we
have determined, and the impact on our interpretation of the data.

\begin{itemize}

\item \emph{Dust depletion}.  Sulphur and oxygen are not expected to have
appreciable depletion on to dust, but boron is mildly refractory
\citep{Howk00}.  Dust depletion would raise the boron abundance above
the gas phase value we have measured, pushing the DLA data points
upwards, and hence even further above the primary line.  Hence, dust
can not explain the possible super-primary boron abundances in the two DLAs.

\item \emph{Ionization effects}.  Ionization has little effect on the oxygen
and sulphur abundances. Corrections for ionization have been modeled
extensively for DLAs in both hard and soft
ionizing radiation fields \citep[e.g. ][]{Vladilo01,Rix07}.
Due to charge exchange reactions, the ionization correction for oxygen
is negligible.  Although photoionization by a stellar radiation field
could lead to an under-estimate of the [S/H] from SII, and push the
DLA data points closer to the $m=1$ line, at the hydrogen column density
of FJ0812+3208 the correction is less than 0.05 dex
\citep{Vladilo01}.    Although boron has not been included in previous
ionization models, the high N(HI) in FJ0812+3208 (logN(HI)=21.35) means
that it is unlikely that a $\sim$ 0.5 dex correction is
required, so we also rule out ionization effects as the reason for
high boron abundances in these two DLAs.

\item \emph{Blending/contamination}.  We can not rule out that the
main features we have identified as boron are in fact due to
contaminating Ly$\alpha$ forest (or other species) lines, 
(e.g. J0058+0115). We also consider the presence of other
weak features within the velocity window over which the integrated
optical depth has been calculated.  For example, the absorption
profiles for FJ0812+3208 in Figure \ref{fig:oheightspec} do show signs
of other absorption features within the velocity window for both
oxygen and boron.

\item \emph{Misplaced continuum}.  For very weak lines, the continuum placement
can be critical. Each spectrum has been visually inspected
to see if the continuum fit seems reasonable and manual
adjustments made if necessary.  Overall, the continuum appears to be well-defined. Section \ref{subsec:continuum} demonstrates that continuum placements can account for a difference in the boron column density by as much as $\sim0.18$ dex in the most extreme case (FJ0812+3208). These continuum offsets can add up to 0.07 dex to the total error budget, thus having a minimal effect to the overall observations. 
\end{itemize}

After excluding corrections due to dust, ionization, continuum errors
and additional components as possible causes for potential super-primary boron
abundances, we now consider the possible physical mechanisms for high
boron abundances.  One intriguing possibility is an enhanced cosmic ray flux that would result in a higher spallation rate of CNO targets to produce the excess boron. Assuming that the cosmic ray flux scales linearly with the boron abundance; J0058+0115 and FJ0812+3208 would have a cosmic ray flux
of $\sim8\times$ and $\sim2\times$ (respectively) the expected flux for primary production of boron in the Milky Way. A high cosmic ray flux could be associated with a line of sight that intersects gas near a supernova remnant, star forming region, or young cluster \citep[for a list of observed sight-lines see][]{Ritchey11}. Such sight-lines through enhanced star forming environments might explain the high N(HI) and high metallicity of the two DLAs with boron detections in our sample. 
\cite{Prochaska03} also note the rare detection of the excited SiII$^{\star}$
in FJ0812+3208 (see Figure \ref{fig:oheightadd}), a species that can be used in conjunction with CII$^{\star}$
to constrain the gas temperature \citep{Howk05}\footnote{This calculation is beyond the scope of the paper. For the interested reader, see \cite{Wolfe03}.}.
Remarkably, J1417+4132 also has a detection of SiII$^{\star}$. To date, SiII$^{\star}$ has only been seen in one other DLA \citep{Kulkarni12}.

As discussed in the Introduction, carbon is also a potential spallation target. Typically 
it is disregarded during cosmic ray spallation as it is less abundant than oxygen.
For example, in the Milky Way, there are typically 6 oxygen nuclei for every carbon, log{N(C)/N(O)}$\sim-0.8$ \citep{Akerman04,Fabbian09}, for $-2.5\leq$[O/H]$\leq-0.5$. This is not
true at [O/H]$\sim-3$ and [O/H]$\sim0$, where [C/O] becomes solar (log{N(C)/N(O)}$\sim-0.3$; 2 oxygen nuclei for every carbon). For our detections in Figure \ref{fig:BS},  we would expect carbon spallation to contribute to boron production slightly. Assuming that there is one carbon atom for every two oxygen as an upper limit, we would expect a shift of $\sim0.18$ dex to the right in Figure \ref{fig:BS} to account for the extra spallation target (as there are $1.5\times$ as many targets). However, this is an overestimate as there are probably more than two oxygen per carbon atom (as [O/H]$<0$) so both detected points would still lie above the primary line. If a galaxy were enriched in carbon relative to oxygen, it is plausible that the boron production could exceed
that expected from primary oxygen-only spallation. Due to the difficulty of measuring 
carbon in DLAs (C II $\lambda1334$ is generally saturated whereas C II 
$\lambda2325$ is too weak to be detected), it is difficult to test whether
our DLAs are indeed carbon rich. However studies of
carbon in DLAs \citep{Pettini08,Penprase10,Cooke11} follow a similar pattern
in the [C/O] to the Milky Way at low metallicities.

\begin{figure}
\begin{center}
\includegraphics[width=0.5\textwidth]{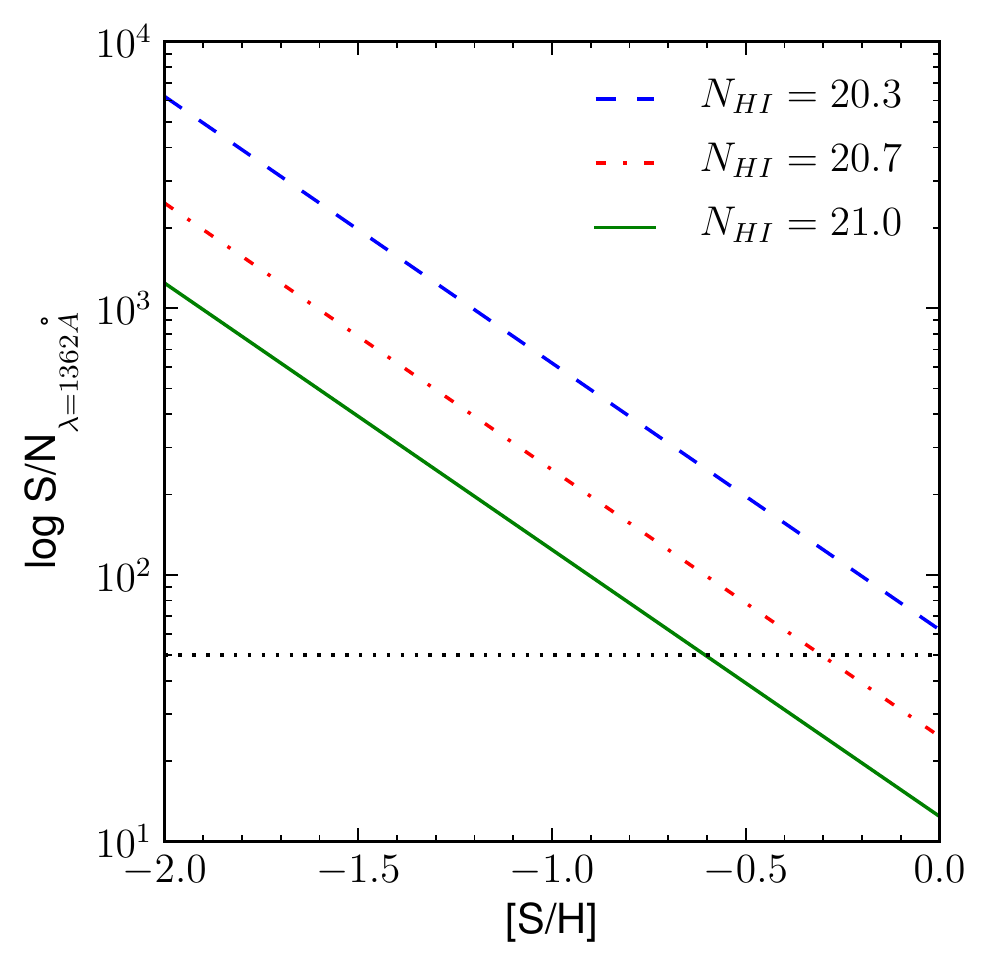}
\caption{The S/N ratio (per pixel) required to observe primary boron is
plotted from Equations \ref{eq:N_lim} and \ref{eq:eqw} for the range
in sulphur values for a fixed redshift of $z_{abs}=2$. A FWHM
corresponding to a $b=10$ \kms{} is assumed as it represents the minimum
detectable line width by HIRES. The black dotted line at S/N of 50
is drawn to represent an achievable S/N ratio.}
\label{fig:snr}
\end{center}
\end{figure}

The results presented in this study demonstrate the possibility of
boron detection in DLAs, but also clearly illustrate its challenges.
In most of the sight-lines of our study, despite the selection of
DLAs with strong metal lines, the upper limits are not deep enough
to constrain boron production.  In order to assess future prospects
for more boron detections, we show in Figure \ref{fig:snr}
the required S/N ratios per pixel at the BII $\lambda1362$ line
as a function of the sulphur abundance based on Equations
\ref{eq:eqw} and \ref{eq:N_lim} for three different hydrogen column
densities, assuming the corresponding FHWM for a $b$-value\footnote{The $b$-value is related to the FWHM by $FWHM= 2\sqrt{ln2} b$} of 10 \kms{}.
Figure \ref{fig:snr} suggests that for a DLA with a reasonably high log N(HI)=21 and high metallicity ([S/H]$>-0.6$), boron that has been synthesized via a primary process could be detected in a spectrum with S/N$\sim50$, which
remains moderately demanding on current 8--10 metre class telescopes. Assuming that the exposure time
varies as the square of the S/N ratio, our sample of DLAs with upper
limits of boron would need exposure times of at least an order of
magnitude greater than those listed in Table \ref{tab:MSDLAs}.
The detection of high redshift boron would be a excellent science
case for the next generation of large optical telescopes with blue
optimized echelle spectrographs, such as the High-Resolution Optical Spectrometer for the Thirty Meter Telescope.

\section{Conclusion}
We have presented 2 detections of boron at $>3\sigma$, 21 upper limits, and one further DLA in which the absorption is suggestive of a boron detection, but formally only significant at 2.97 sigma. The results hint at an excess of boron relative 
to predictions from primary production in the two cases where we
have positive detections; albeit they are still consistent with the primary production mechanism. Although they are presented, it is unclear whether detections 
are truly real. We rule out that the potential
super-primary boron abundances are due to dust or ionization effects
and discuss possible physical reasons for the overabundance. Higher
boron abundances might be due to higher cosmic ray fluxes in these two
DLAs, relative to the Galactic value.  This study also has shown that
sulphur can be used in place of oxygen for studies of boron. This is
useful as oxygen abundances are often unreliable and difficult to
obtain in the Milky Way stars, ISM, and DLAs. Figure \ref{fig:snr}
demonstrates that for high S/N ratios of $\sim50$, DLAs with high
metallicites ([S/H]$>-$0.6 dex) and high HI columns
(logN(HI)$\sim21$) are needed to test whether DLAs do in fact exhibit
higher boron abundances relative to primary production.

\section*{Acknowledgments}

We would like to thank the anonymous referee for their useful comments to ensure the clarity of this work. We gratefully acknowledge the efforts of M. Neeleman in reducing some of the spectra presented here and for his helpful comments on this manuscript. JXP acknowledges support from NSF grant AST-1109447. The authors wish to recognize and acknowledge the very significant cultural role and reverence that the summit of Mauna Kea has always had within the indigenous Hawaiian community. We are most fortunate to have the opportunity to conduct observations from this mountain.

\bibliography{bref}
\onecolumn
\begin{longtable}{lcccc}

\caption{Literature boron, oxygen, and sulphur abundances}
\label{tab:stellarBSO}\\
\hline
ID	 &[O/H]	&[S/H]	&[B/H]	&References\\
\hline
\endfirsthead

\multicolumn{2}{l}{{\bfseries \tablename\ \thetable{} -- continued from previous page}}\\
\hline
ID	 &[O/H]	&[S/H]	&[B/H]	&References\\
\hline
\endhead
\hline \multicolumn{2}{l}{\bfseries Continued on next page}\\
\hline
\endfoot
\endlastfoot
\\
\multicolumn{5}{c}{\textbf{Stars}}\\
\hline
HD19994	 	&$0.25\pm0.06$	&$0.11\pm0.10$&	$0.01\pm0.20$	&1,2\\
HD159332	&$-0.07\pm0.06$&	\nodata	&$-0.39\pm0.20$	&1,2\\
HD5015		&$0.07\pm0.06$	&$-0.01\pm0.10$&	$-0.29\pm0.20$	&1,2\\
HD216385	&$0.04\pm0.06$&	$-0.14\pm0.10$&	$-0.39\pm0.20$	&1,2\\
HD185395	&$0.07\pm0.06$&	$-0.19\pm0.10$&	$-0.19\pm0.20$	&1,2\\
HD184499	&$-0.15\pm0.06$&	\nodata	&$-0.69\pm0.20$	&1,2\\
HD210027	&$-0.04\pm0.06$&	\nodata	&$-0.19\pm0.20$	&1,2\\
HD128167	&$-0.06\pm0.06$&	$-0.33\pm0.10$&	$-0.49\pm0.20$	&1,2\\
HD82328		&$0.07\pm0.06$	&$-0.25\pm0.10$	&$-0.19\pm0.20$	&1,2\\
HD4813		&$0.09\pm0.06$	&$-0.08\pm0.10$	&$-0.09\pm0.20$	&1,2\\
HD28033		&$0.20\pm0.06$	&\nodata	&$-0.19\pm0.20$	&1,2\\
NGC346$-$637	&$-0.69\pm0.2$&	\nodata&$<-1.19$	&3\\
AV304	&$-0.49\pm0.2$	&\nodata&$<-1.19$	&3\\
HD194598	&$-0.69\pm0.10$	&$-0.76\pm0.15$	&$-1.36\pm0.14$	&4,5\\
HD94028		&$-0.87\pm0.10$	&$-1.22\pm0.15$	&$-1.48\pm0.16$	&4,5\\
BD+23\degree{}3130	&$-1.58\pm0.10$	&\nodata	&$-2.88\pm0.25$	&4\\
HD76932	&$-0.35\pm0.10$	&$-0.68\pm0.15$	&$-0.72\pm0.14$	&4,5\\
HD201891	&$-0.56\pm0.10$	&$-0.81\pm0.15$	&$-1.03\pm0.15$	&4,5\\
HD19445		&$-1.32\pm0.10$	&$-1.42\pm0.15$	&$-2.12\pm0.19$	&4,5\\
HD84937		&$-1.61\pm0.10$	&$-1.68\pm0.15$	&$<-2.15$	&4,5\\
BD+26\degree{}3578	&$-1.65\pm0.10$	&\nodata	&$<-2.35$	&4\\
HD160617	&$-1.30\pm0.10$	&$-1.35\pm0.15$	&$-2.18\pm0.20$	&4,5\\
HD184499	&$-0.19\pm0.10$	&\nodata	&$-0.56\pm0.17$	&4\\
HD64090		&$-1.00\pm0.10$	&\nodata	&$-1.50\pm0.13$	&4\\
BD+03\degree{}740	&$-2.08\pm0.10$	&\nodata	&$<-2.13$	&4\\
BD$-$13\degree{}3442	&$-2.14\pm0.10$	&$-2.26\pm0.15$	&$<-2.35$	&4,5\\
HD106516	&$-0.22\pm0.10$	&$-0.34\pm0.15$	&$-1.36\pm0.15$	&4,5\\
HD140283	&$-1.69\pm0.10$	&$-2.06\pm0.15$	&$-2.60\pm0.22$	&4,5\\
HD221377&$-0.55\pm0.10$	&\nodata	&$-1.30\pm0.19$	&4\\
HD37744	&$-0.23\pm0.20$	&\nodata	&$-0.29\pm0.10$	&6\\
HD44743	&$0.09\pm0.20$	&\nodata	&$0.01\pm0.10$	&6\\
HD36959	&$-0.08\pm0.20$	&$0.01\pm0.14$	&$-0.29\pm0.10$&	6,7\\
HD36629	&$-0.37\pm0.20$	&\nodata	&$-0.29\pm0.10$	&6\\
HD36351	&$-0.17\pm0.20$	&\nodata	&$-0.19\pm0.10$	&6\\
HD214993	&$0.14\pm0.20$&	\nodata	&$-0.49\pm0.10$	&6\\
HD216916	&$-0.08\pm0.20$&	\nodata	&$-0.48\pm0.10$	&6\\
HD35337	&$-0.14\pm0.20$	&\nodata	&$-0.69\pm0.10$	&6\\
HD37356	&$-0.25\pm0.20$	&$-0.02\pm0.14$	&$-0.29\pm0.10$	&6,7\\
HD35039	&$-0.35\pm0.20$	&$0.03\pm0.15$	&$0.13\pm0.10$	&6,7\\
HD29248	&$-0.02\pm0.20$	&\nodata	&$-0.29\pm0.10$	&6\\
BD+56\degree{}576	&$-0.35\pm0.20$&	\nodata	&$-0.54\pm0.10$	&6\\
HD34816	&$-0.02\pm0.20$	&\nodata	&$-0.49\pm0.10$	&6\\

LP815$-$43& 	$-1.86\pm0.15$ & 		$-2.49\pm0.22$& 	\nodata& 	8,9\\
CS22873$-$055& 	$-2.42\pm0.25$ & 		$-2.53\pm0.10$& 	\nodata& 	10,11\\
CS30325$-$094& 	$-2.53\pm0.25$ & 		$-2.90\pm0.18$& 	\nodata& 	10,11\\
HD179626& 	$-0.46\pm0.03$ & 		$-0.90\pm0.22$& 	\nodata& 	8,12\\
CS22948$-$066& 	$-2.20\pm0.25$ & 		$-2.83\pm0.14$& 	\nodata& 	10,11\\
G64$-$12& 	$-2.24\pm0.15$ & 		$-3.00\pm0.22$& 	\nodata& 	8,9\\
G18-39& 	$-0.87\pm0.04$ & 		$-1.15\pm0.22$& 	\nodata& 	8,12\\
HD148816& 	$-0.24\pm0.03$ & 		$-0.44\pm0.15$& 	\nodata& 	5,12\\
CS29518$-$051& 	$-1.84\pm0.25$ & 		$-2.58\pm0.10$& 	\nodata& 	10,11\\
CS22186$-$025& 	$-2.36\pm0.25$ & 		$-2.66\pm0.10$& 	\nodata& 	10,11\\
LP635$-$14& 	$-2.00\pm0.12$ & 		$-2.20\pm0.22$& 	\nodata& 	8,9\\
CS22896$-$154& 	$-1.70\pm0.25$ & 		$-2.52\pm0.08$& 	\nodata& 	10,11\\
HD2796& 	$-1.92\pm0.25$ & 		$-2.11\pm0.10$& 	\nodata& 	10,11\\
G11$-$44& 	$-1.63\pm0.15$ & 		$-1.71\pm0.22$& 	\nodata& 	8,9\\
HD106516& 	$-0.16\pm0.03$ & 		$-0.34\pm0.15$& 	\nodata& 	5,12\\
HD76932& 	$-0.37\pm0.03$ & 		$-0.68\pm0.15$& 	\nodata& 	5,12\\
CS29491$-$053& 	$-2.23\pm0.25$ & 		$-2.65\pm0.10$& 	\nodata& 	10,11\\
CS22891$-$209& 	$-2.47\pm0.25$ & 		$-2.88\pm0.15$& 	\nodata& 	10,11\\
G53$-$41& 	$-0.99\pm0.05$ & 		$-1.04\pm0.22$& 	\nodata& 	8,12\\
HD194598& 	$-0.67\pm0.03$ & 		$-0.97\pm0.22$& 	\nodata& 	8,12\\
LP651$-$4& 	$-2.04\pm0.14$ & 		$-2.47\pm0.22$& 	\nodata& 	8,9\\
CS22966$-$057& 	$-1.58\pm0.25$ & 		$-2.45\pm0.10$& 	\nodata& 	10,11\\
G64$-$37& 	$-2.32\pm0.14$ & 		$-2.93\pm0.22$& 	\nodata& 	8,9\\
BD$-$13\degree{}3442& 	$-2.15\pm0.15$ & 		$-2.41\pm0.22$& 	\nodata& 	8,9\\
BD+17\degree{}3248& 	$-1.33\pm0.25$ & 		$-1.94\pm0.10$& 	\nodata& 	10,11\\
CS31082$-$001& 	$-2.26\pm0.25$ & 		$-2.54\pm0.10$& 	\nodata& 	10,11\\
BD$-$18\degree{}5550& 	$-2.59\pm0.25$ & 		$-2.67\pm0.12$& 	\nodata& 	10,11\\
HD122563& 	$-2.15\pm0.25$ & 		$-2.39\pm0.10$& 	\nodata& 	10,11\\
HD186478& 	$-1.79\pm0.25$ & 		$-2.25\pm0.08$& 	\nodata& 	10,11\\
CS22953$-$003& 	$-2.04\pm0.25$ & 		$-2.68\pm0.10$& 	\nodata& 	10,11\\
HD193901& 	$-0.64\pm0.03$ & 		$-0.88\pm0.22$& 	\nodata& 	8,12\\
CS22956$-$050& 	$-2.16\pm0.25$ & 		$-2.91\pm0.20$& 	\nodata& 	10,11\\
CS22892$-$052& 	$-2.51\pm0.25$ & 		$-2.81\pm0.10$& 	\nodata& 	10,11\\
HD9091& 	$-0.04\pm0.15$ & 		$-0.32\pm0.15$& 	\nodata& 	13\\
HD20427& 	$-0.12\pm0.15$ & 		$-0.39\pm0.15$& 	\nodata& 	13\\
HD112887& 	$0.02\pm0.15$ & 		$-0.26\pm0.15$& 	\nodata& 	13\\
HD2663& 	$-0.11\pm0.15$ & 		$-0.37\pm0.15$& 	\nodata& 	13\\
HD209858& 	$0.00\pm0.15$ & 		$-0.23\pm0.15$& 	\nodata& 	13\\
HD7228& 	$0.13\pm0.15$ & 		$-0.07\pm0.15$& 	\nodata& 	13\\
HD157467& 	$0.30\pm0.15$ & 		$0.21\pm0.15$& 	\nodata& 	13\\
HD218172& 	$0.07\pm0.15$ & 		$-0.15\pm0.15$& 	\nodata& 	13\\
HD133641& 	$-0.06\pm0.15$ & 		$-0.34\pm0.15$& 	\nodata& 	13\\
HD210457& 	$0.04\pm0.15$ & 		$-0.22\pm0.15$& 	\nodata& 	13\\
HD222155& 	$0.09\pm0.15$ & 		$-0.18\pm0.15$& 	\nodata& 	13\\
HD103891& 	$-0.06\pm0.15$ & 		$-0.21\pm0.15$& 	\nodata& 	13\\
HD42618& 	$0.02\pm0.15$ & 		$-0.15\pm0.15$& 	\nodata& 	13\\
HD80218& 	$0.10\pm0.15$ & 		$-0.19\pm0.15$& 	\nodata& 	13\\
HD109303& 	$0.11\pm0.15$ & 		$-0.36\pm0.15$& 	\nodata& 	13\\
HD198390& 	$0.02\pm0.15$ & 		$-0.30\pm0.15$& 	\nodata& 	13\\
HD59360& 	$0.13\pm0.15$ & 		$-0.16\pm0.15$& 	\nodata& 	13\\
HD110989& 	$0.11\pm0.15$ & 		$-0.25\pm0.15$& 	\nodata& 	13\\
HD171620& 	$-0.16\pm0.15$ & 		$-0.39\pm0.15$& 	\nodata& 	13\\
HD121560& 	$-0.13\pm0.15$ & 		$-0.38\pm0.15$& 	\nodata& 	13\\
HD5065& 	$0.12\pm0.15$ & 		$-0.12\pm0.15$& 	\nodata& 	13\\
HD155646& 	$0.27\pm0.15$ & 		$-0.03\pm0.15$& 	\nodata& 	13\\
HD11007& 	$-0.09\pm0.15$ & 		$-0.29\pm0.15$& 	\nodata& 	13\\
HD153240& 	$0.15\pm0.15$ & 		$-0.05\pm0.15$& 	\nodata& 	13\\
HD191672& 	$-0.16\pm0.15$ & 		$-0.38\pm0.15$& 	\nodata& 	13\\
HD73400& 	$0.00\pm0.15$ & 		$-0.17\pm0.15$& 	\nodata& 	13\\
HD63333& 	$-0.06\pm0.15$ & 		$-0.33\pm0.15$& 	\nodata& 	13\\
HD15398& 	$0.23\pm0.15$ & 		$0.02\pm0.15$& 	\nodata& 	13\\
HD214576& 	$-0.18\pm0.15$ & 		$-0.44\pm0.15$& 	\nodata& 	13\\
HD107038& 	$-0.21\pm0.15$ & 		$-0.33\pm0.15$& 	\nodata& 	13\\
HD22718& 	$0.15\pm0.15$ & 		$-0.13\pm0.15$& 	\nodata& 	13\\
HD182758& 	$-0.14\pm0.15$ & 		$-0.49\pm0.15$& 	\nodata& 	13\\
HD186379& 	$-0.11\pm0.15$ & 		$-0.33\pm0.15$& 	\nodata& 	13\\
HD201444& 	$-0.10\pm0.15$ & 		$-0.50\pm0.15$& 	\nodata& 	13\\
HD216106& 	$0.10\pm0.15$ & 		$-0.24\pm0.15$& 	\nodata& 	13\\
HD159333& 	$0.12\pm0.15$ & 		$-0.18\pm0.15$& 	\nodata& 	13\\
HD153668& 	$0.04\pm0.15$ & 		$-0.17\pm0.15$& 	\nodata& 	13\\
HD153627& 	$-0.04\pm0.15$ & 		$-0.30\pm0.15$& 	\nodata& 	13\\
HD76349& 	$-0.16\pm0.15$ & 		$-0.39\pm0.15$& 	\nodata& 	13\\
HD77408& 	$-0.09\pm0.15$ & 		$-0.21\pm0.15$& 	\nodata& 	13\\
HD26421& 	$-0.03\pm0.15$ & 		$-0.32\pm0.15$& 	\nodata& 	13\\
HD330& 	$0.08\pm0.15$ & 		$-0.20\pm0.15$& 	\nodata& 	13\\
HD210985& 	$-0.21\pm0.15$ & 		$-0.46\pm0.15$& 	\nodata& 	13\\
HD200580& 	$-0.37\pm0.15$ & 		$-0.53\pm0.15$& 	\nodata& 	13\\
HD22255& 	$-0.01\pm0.15$ & 		$-0.22\pm0.15$& 	\nodata& 	13\\
HD63332& 	$0.21\pm0.15$ & 		$0.06\pm0.15$& 	\nodata& 	13\\
HD88446& 	$0.06\pm0.15$ & 		$-0.26\pm0.15$& 	\nodata& 	13\\
HD69897& 	$-0.05\pm0.15$ & 		$-0.20\pm0.15$& 	\nodata& 	13\\
HD199085& 	$0.03\pm0.15$ & 		$-0.10\pm0.15$& 	\nodata& 	13\\
HD140750& 	$0.14\pm0.15$ & 		$-0.26\pm0.15$& 	\nodata& 	13\\
HD100446& 	$-0.16\pm0.15$ & 		$-0.43\pm0.15$& 	\nodata& 	13\\
HD149576& 	$0.16\pm0.15$ & 		$-0.14\pm0.15$& 	\nodata& 	13\\
HD126053& 	$-0.16\pm0.15$ & 		$-0.34\pm0.15$& 	\nodata& 	13\\
HD87838& 	$-0.21\pm0.15$ & 		$-0.35\pm0.15$& 	\nodata& 	13\\
HD219497& 	$-0.07\pm0.15$ & 		$-0.42\pm0.15$& 	\nodata& 	13\\
HD52711& 	$0.05\pm0.15$ & 		$-0.16\pm0.15$& 	\nodata& 	13\\
HD210718& 	$-0.16\pm0.15$ & 		$-0.27\pm0.15$& 	\nodata& 	13\\
HD140324& 	$-0.01\pm0.15$ & 		$-0.28\pm0.15$& 	\nodata& 	13\\
HD24421& 	$-0.07\pm0.15$ & 		$-0.35\pm0.15$& 	\nodata& 	13\\
HD136925& 	$0.11\pm0.15$ & 		$-0.18\pm0.15$& 	\nodata& 	13\\
HD101& 	$-0.05\pm0.15$ & 		$-0.24\pm0.15$& 	\nodata& 	13\\
HD218059& 	$-0.02\pm0.15$ & 		$-0.24\pm0.15$& 	\nodata& 	13\\
HD152986& 	$0.06\pm0.15$ & 		$-0.12\pm0.15$& 	\nodata& 	13\\
HD22521& 	$0.03\pm0.15$ & 		$-0.21\pm0.15$& 	\nodata& 	13\\
HD45067& 	$0.07\pm0.15$ & 		$-0.09\pm0.15$& 	\nodata& 	13\\
HD41640& 	$-0.30\pm0.15$ & 		$-0.50\pm0.15$& 	\nodata& 	13\\
HD204712& 	$-0.04\pm0.15$ & 		$-0.38\pm0.15$& 	\nodata& 	13\\
HD224233& 	$0.07\pm0.15$ & 		$-0.12\pm0.15$& 	\nodata& 	13\\
HD9670& 	$-0.07\pm0.15$ & 		$-0.24\pm0.15$& 	\nodata& 	13\\
HD108134& 	$-0.06\pm0.15$ & 		$-0.32\pm0.15$& 	\nodata& 	13\\
HD131599& 	$-0.07\pm0.15$ & 		$-0.40\pm0.15$& 	\nodata& 	13\\
HD127667& 	$-0.13\pm0.15$ & 		$-0.34\pm0.15$& 	\nodata& 	13\\
HD6840& 	$-0.10\pm0.15$ & 		$-0.34\pm0.15$& 	\nodata& 	13\\
HD167588& 	$-0.08\pm0.15$ & 		$-0.33\pm0.15$& 	\nodata& 	13\\
HD198089& 	$0.00\pm0.15$ & 		$-0.24\pm0.15$& 	\nodata& 	13\\
HD5750& 	$-0.19\pm0.15$ & 		$-0.31\pm0.15$& 	\nodata& 	13\\
HD193664& 	$-0.02\pm0.15$ & 		$-0.17\pm0.15$& 	\nodata& 	13\\
HD218637& 	$-0.13\pm0.15$ & 		$-0.22\pm0.15$& 	\nodata& 	13\\
HD221356& 	$-0.15\pm0.15$ & 		$-0.25\pm0.15$& 	\nodata& 	13\\
HD3454& 	$-0.22\pm0.15$ & 		$-0.51\pm0.15$& 	\nodata& 	13\\
HD16067& 	$0.23\pm0.15$ & 		$-0.04\pm0.15$& 	\nodata& 	13\\
HD91638& 	$-0.03\pm0.15$ & 		$-0.24\pm0.15$& 	\nodata& 	13\\
HD146946& 	$-0.07\pm0.15$ & 		$-0.33\pm0.15$& 	\nodata& 	13\\
HD204559& 	$-0.04\pm0.15$ & 		$-0.32\pm0.15$& 	\nodata& 	13\\
HD94835& 	$0.24\pm0.15$ & 		$0.12\pm0.15$& 	\nodata& 	13\\
HD198109& 	$-0.15\pm0.15$ & 		$-0.41\pm0.15$& 	\nodata& 	13\\
HD131039& 	$0.18\pm0.15$ & 		$-0.14\pm0.15$& 	\nodata& 	13\\
HD36066& 	$0.16\pm0.15$ & 		$0.01\pm0.15$& 	\nodata& 	13\\
HD5494& 	$0.18\pm0.15$ & 		$-0.06\pm0.15$& 	\nodata& 	13\\
HD210923& 	$0.12\pm0.15$ & 		$-0.17\pm0.15$& 	\nodata& 	13\\
HD212858& 	$-0.10\pm0.15$ & 		$-0.39\pm0.15$& 	\nodata& 	13\\
HD157466& 	$-0.17\pm0.15$ & 		$-0.39\pm0.15$& 	\nodata& 	13\\
HD220908& 	$0.15\pm0.15$ & 		$-0.10\pm0.15$& 	\nodata& 	13\\
HD3532& 	$0.02\pm0.15$ & 		$-0.28\pm0.15$& 	\nodata& 	13\\
HD109154& 	$0.01\pm0.15$ & 		$-0.32\pm0.15$& 	\nodata& 	13\\
HD174160& 	$0.12\pm0.15$ & 		$-0.06\pm0.15$& 	\nodata& 	13\\
HD102618& 	$0.03\pm0.15$ & 		$-0.26\pm0.15$& 	\nodata& 	13\\
HD86884& 	$-0.01\pm0.15$ & 		$-0.24\pm0.15$& 	\nodata& 	13\\
HD223436& 	$0.21\pm0.15$ & 		$0.03\pm0.15$& 	\nodata& 	13\\
HD15029& 	$-0.10\pm0.15$ & 		$-0.28\pm0.15$& 	\nodata& 	13\\
HD6312& 	$-0.04\pm0.15$ & 		$-0.28\pm0.15$& 	\nodata& 	13\\
HD3440& 	$-0.03\pm0.15$ & 		$-0.30\pm0.15$& 	\nodata& 	13\\
HD214111& 	$0.20\pm0.15$ & 		$-0.05\pm0.15$& 	\nodata& 	13\\
HD100067& 	$-0.11\pm0.15$ & 		$-0.27\pm0.15$& 	\nodata& 	13\\
HD118687& 	$-0.11\pm0.15$ & 		$-0.38\pm0.15$& 	\nodata& 	13\\
HD218470& 	$0.23\pm0.15$ & 		$-0.07\pm0.15$& 	\nodata& 	13\\
HD102080& 	$-0.14\pm0.15$ & 		$-0.32\pm0.15$& 	\nodata& 	13\\
HD11592& 	$-0.05\pm0.15$ & 		$-0.27\pm0.15$& 	\nodata& 	13\\
HD220842& 	$0.09\pm0.15$ & 		$-0.24\pm0.15$& 	\nodata& 	13\\
HD11045& 	$0.07\pm0.15$ & 		$-0.25\pm0.15$& 	\nodata& 	13\\
HD106510& 	$-0.16\pm0.15$ & 		$-0.39\pm0.15$& 	\nodata& 	13\\
HD186408& 	$0.21\pm0.15$ & 		$0.06\pm0.15$& 	\nodata& 	13\\
HD204306& 	$-0.13\pm0.15$ & 		$-0.52\pm0.15$& 	\nodata& 	13\\
HD201835& 	$-0.04\pm0.15$ & 		$-0.28\pm0.15$& 	\nodata& 	13\\
HD3079& 	$0.08\pm0.15$ & 		$-0.16\pm0.15$& 	\nodata& 	13\\
HD210640& 	$0.01\pm0.15$ & 		$-0.28\pm0.15$& 	\nodata& 	13\\
HD163363& 	$0.17\pm0.15$ & 		$0.01\pm0.15$& 	\nodata& 	13\\
HD152449& 	$0.16\pm0.15$ & 		$-0.03\pm0.15$& 	\nodata& 	13\\
HD86560& 	$-0.08\pm0.15$ & 		$-0.37\pm0.15$& 	\nodata& 	13\\
HD101716& 	$0.20\pm0.15$ & 		$-0.11\pm0.15$& 	\nodata& 	13\\
HD216385& 	$0.02\pm0.15$ & 		$-0.15\pm0.15$& 	\nodata& 	13\\
HD89010& 	$0.23\pm0.15$ & 		$0.02\pm0.15$& 	\nodata& 	13\\
HD101676& 	$-0.12\pm0.15$ & 		$-0.37\pm0.15$& 	\nodata& 	13\\
HD139457& 	$-0.11\pm0.15$ & 		$-0.41\pm0.15$& 	\nodata& 	13\\
HD190681& 	$0.17\pm0.15$ & 		$-0.02\pm0.15$& 	\nodata& 	13\\
HD21922& 	$-0.07\pm0.15$ & 		$-0.41\pm0.15$& 	\nodata& 	13\\
HD99126& 	$0.07\pm0.15$ & 		$-0.10\pm0.15$& 	\nodata& 	13\\
HD20717& 	$-0.11\pm0.15$ & 		$-0.27\pm0.15$& 	\nodata& 	13\\
HD6250& 	$0.06\pm0.15$ & 		$-0.13\pm0.15$& 	\nodata& 	13\\
HD77134& 	$-0.08\pm0.15$ & 		$-0.26\pm0.15$& 	\nodata& 	13\\
HD97037& 	$0.03\pm0.15$ & 		$-0.12\pm0.15$& 	\nodata& 	13\\
HD130253& 	$0.06\pm0.15$ & 		$-0.16\pm0.15$& 	\nodata& 	13\\
HD124819& 	$0.08\pm0.15$ & 		$-0.22\pm0.15$& 	\nodata& 	13\\
HD70& 	$-0.06\pm0.15$ & 		$-0.27\pm0.15$& 	\nodata& 	13\\
HD85902& 	$-0.17\pm0.15$ & 		$-0.50\pm0.15$& 	\nodata& 	13\\
HD219476& 	$-0.17\pm0.15$ & 		$-0.46\pm0.15$& 	\nodata& 	13\\
HD217877& 	$0.04\pm0.15$ & 		$-0.14\pm0.15$& 	\nodata& 	13\\
HD223854& 	$-0.24\pm0.15$ & 		$-0.47\pm0.15$& 	\nodata& 	13\\
HD171886& 	$0.00\pm0.15$ & 		$-0.28\pm0.15$& 	\nodata& 	13\\
HD153& 	$0.29\pm0.15$ & 		$0.00\pm0.15$& 	\nodata& 	13\\
HD3894& 	$-0.20\pm0.15$ & 		$-0.35\pm0.15$& 	\nodata& 	13\\
HD214435& 	$-0.07\pm0.15$ & 		$-0.28\pm0.15$& 	\nodata& 	13\\
HD202884& 	$-0.03\pm0.15$ & 		$-0.22\pm0.15$& 	\nodata& 	13\\
HD54182& 	$0.18\pm0.15$ & 		$-0.16\pm0.15$& 	\nodata& 	13\\
HD112756& 	$-0.13\pm0.15$ & 		$-0.31\pm0.15$& 	\nodata& 	13\\
HD169359& 	$-0.01\pm0.15$ & 		$-0.23\pm0.15$& 	\nodata& 	13\\
HD225239& 	$-0.10\pm0.15$ & 		$-0.44\pm0.15$& 	\nodata& 	13\\
HD219983& 	$0.02\pm0.15$ & 		$-0.17\pm0.15$& 	\nodata& 	13\\
HD76272& 	$-0.09\pm0.15$ & 		$-0.35\pm0.15$& 	\nodata& 	13\\
HD219306& 	$-0.01\pm0.15$ & 		$-0.25\pm0.15$& 	\nodata& 	13\\
HD14877& 	$-0.12\pm0.15$ & 		$-0.36\pm0.15$& 	\nodata& 	13\\
HD201490& 	$-0.04\pm0.15$ & 		$-0.20\pm0.15$& 	\nodata& 	13\\
HD216631& 	$-0.14\pm0.15$ & 		$-0.41\pm0.15$& 	\nodata& 	13\\
HD148049& 	$-0.08\pm0.15$ & 		$-0.31\pm0.15$& 	\nodata& 	13\\
HD214557& 	$0.21\pm0.15$ & 		$0.02\pm0.15$& 	\nodata& 	13\\
HD201639& 	$-0.32\pm0.15$ & 		$-0.45\pm0.15$& 	\nodata& 	13\\
HD176796& 	$0.00\pm0.15$ & 		$-0.32\pm0.15$& 	\nodata& 	13\\
HD195200& 	$0.05\pm0.15$ & 		$-0.19\pm0.15$& 	\nodata& 	13\\
HD23438& 	$-0.02\pm0.15$ & 		$-0.29\pm0.15$& 	\nodata& 	13\\
HD90878& 	$0.15\pm0.15$ & 		$-0.19\pm0.15$& 	\nodata& 	13\\
HD94012& 	$-0.14\pm0.15$ & 		$-0.43\pm0.15$& 	\nodata& 	13\\
HD27816& 	$-0.25\pm0.15$ & 		$-0.47\pm0.15$& 	\nodata& 	13\\
HD191649& 	$0.06\pm0.15$ & 		$-0.18\pm0.15$& 	\nodata& 	13\\
HD215442& 	$0.28\pm0.15$ & 		$-0.11\pm0.15$& 	\nodata& 	13\\
HD101472& 	$0.11\pm0.15$ & 		$-0.15\pm0.15$& 	\nodata& 	13\\
HD71148& 	$0.07\pm0.15$ & 		$-0.07\pm0.15$& 	\nodata& 	13\\
HD223583& 	$-0.14\pm0.15$ & 		$-0.41\pm0.15$& 	\nodata& 	13\\
HD8671& 	$0.10\pm0.15$ & 		$-0.15\pm0.15$& 	\nodata& 	13\\
HD6834& 	$-0.30\pm0.15$ & 		$-0.53\pm0.15$& 	\nodata& 	13\\
HD36909& 	$0.08\pm0.15$ & 		$-0.18\pm0.15$& 	\nodata& 	13\\
HD99984& 	$-0.03\pm0.15$ & 		$-0.30\pm0.15$& 	\nodata& 	13\\
HD912& 	$0.17\pm0.15$ & 		$-0.18\pm0.15$& 	\nodata& 	13\\
HD209320& 	$0.16\pm0.15$ & 		$-0.10\pm0.15$& 	\nodata& 	13\\
HD36667& 	$-0.06\pm0.15$ & 		$-0.35\pm0.15$& 	\nodata& 	13\\
HD156635& 	$0.16\pm0.15$ & 		$-0.03\pm0.15$& 	\nodata& 	13\\
HD194497& 	$-0.07\pm0.15$ & 		$-0.26\pm0.15$& 	\nodata& 	13\\
HD192145& 	$-0.04\pm0.15$ & 		$-0.28\pm0.15$& 	\nodata& 	13\\
HD160078& 	$0.36\pm0.15$ & 		$-0.03\pm0.15$& 	\nodata& 	13\\
HD99233& 	$-0.24\pm0.15$ & 		$-0.54\pm0.15$& 	\nodata& 	13\\
HD206860& 	$0.10\pm0.15$ & 		$-0.10\pm0.15$& 	\nodata& 	13\\
HD145937& 	$-0.11\pm0.15$ & 		$-0.52\pm0.15$& 	\nodata& 	13\\
\hline
\\
\multicolumn{5}{c}{\textbf{DLAs}}\\
\hline
J1340+1106	& $-1.65\pm0.09$	&$-1.85\pm0.07$	&\nodata	&14\\
Q2059$-$360	& $-1.58\pm0.10$	&$-1.72\pm0.09$	&\nodata	&15,16\\
Q0841+12	& $-1.32\pm0.14$	&$-1.45\pm0.13$	&\nodata	&16,17\\
HE2243$-$6031	& $-1.15\pm0.20$	&$-0.94\pm0.03$	&\nodata	&18\\
Q1337+113	& $-1.95\pm0.12$	&$-1.82\pm0.08$	&\nodata	&16,19\\
\hline
\\
\multicolumn{5}{c}{\textbf{ISM}}\\
\hline
HD88115 & $0.03\pm0.16$ & \nodata & $-0.28\pm0.18$ & 20\\
HD92554 & $0.03\pm0.14$ & \nodata & $-0.35\pm0.16$ & 20\\
HD99890 & $0.25\pm0.17$ & \nodata & $-0.41\pm0.25$ & 20\\
HD104705 & $0.01\pm0.08$ & \nodata & $-0.42\pm0.10$ & 20\\
HD121968 & $0.02\pm0.11$ & \nodata & $-0.51\pm0.21$ & 20\\
HD177989 & $0.05\pm0.09$ & \nodata & $-0.53\pm0.12$ & 20\\
\hline
\end{longtable}
\textsc{References} --
		(1) \cite{Smith01}.
		(2) \cite{Takeda07}.
		(3) \cite{Brooks02} and references therein. 
		(4) \cite{Tan10}. 
 		(5) \cite{Caffau05}.
		(6) \cite{Venn02} and references therein.
		(7) \cite{Daflon09}. 
		(8) \cite{Nissen07}..
		(9) \cite{Rich09}. 
		(10) \cite{Spite05}.		
		(11) \cite{Spite11}.
		(12) \cite{Ramirez12}.
		(13) \cite{Reddy03}
		(14) \cite{Cooke11}.
		(15) \cite{Srianand05}.
		(16) \cite{Petitjean08}.
		(17) \cite{Dessauges07}.
		(18) \cite{Lopez02}.
		(19) \cite{Prochaska07}.
		(20) \cite{Ritchey11}.

\twocolumn
\end{document}